\documentclass[pra,aps,showpacs,groupedaddress,superscriptaddress,twocolumn,toc=flat,biblatex,footinbib]{revtex4-1}
\usepackage{natbib}
\usepackage[table]{xcolor}
\usepackage[colorlinks=false]{hyperref}
\usepackage[utf8]{inputenc}
\usepackage{color}
\usepackage{bbm} 
\usepackage[english]{babel}
\usepackage{multirow}
\usepackage{amsfonts,amsmath,amssymb,stmaryrd}

\usepackage{graphicx}
\usepackage{subfigure}  

\usepackage{epsfig}
\usepackage{mathrsfs}
\usepackage{verbatim}
\usepackage{centernot}
\usepackage{ulem}
\usepackage{longtable}
\usepackage{nicefrac}
\usepackage[framemethod=tikz]{mdframed}
\usepackage{siunitx}
\usepackage[nolist]{acronym}

\usepackage{soul}
\usepackage{array}

\usepackage{cancel,ifthen}
\newcommand{\cmnt}[2][NoInPuT]{\ifthenelse{\equal{#1}{NoInPuT}}{}{{\color{red}\sout{#1}}} {\color{blue} #2}}

\usepackage{bm}	
\renewcommand{\vec}[1]{\bm{#1}}

\LTcapwidth=\textwidth

\begin{document}

\normalem	

\title{N\'eel-Vector Switching and THz Spin-Wave Excitation in Mn$_{2}$Au due to Femtosecond Spin-Transfer Torques}

\author{Markus Weißenhofer}
\email[]{markus.weissenhofer@fu-berlin.de}
 \affiliation{Department of Physics and Astronomy, Uppsala University, P.\ O.\ Box 516, S-751 20 Uppsala, Sweden}
\affiliation{Department of Physics, Freie Universit{\"a}t Berlin, Arnimallee 14, D-14195 Berlin, Germany}

\author{Francesco Foggetti}
\affiliation{Department of Physics and Astronomy, Uppsala University, P.\ O.\ Box 516, S-751 20 Uppsala, Sweden}

\author{Ulrich Nowak}
\affiliation{Department of Physics, University of Konstanz,
D-78457 Konstanz, Germany}

\author{Peter M. Oppeneer}
\affiliation{Department of Physics and Astronomy, Uppsala University, P.\ O.\ Box 516, S-751 20 Uppsala, Sweden}


\date{\today}

\begin{abstract}
Efficient and fast manipulation of antiferromagnets has to date remained a challenging task, hindering their application in spintronic devices. For ultrafast operation of such devices, it is highly desirable to be able to control the antiferromagnetic order within picoseconds - a timescale that is difficult to achieve with electrical circuits. Here, we demonstrate that bursts of spin-polarized hot-electron currents emerging due to laser-induced ultrafast demagnetization are able to efficiently excite spin dynamics in antiferromagnetic Mn$_2$Au by exerting a spin-transfer torque on femtosecond timescales. We combine quantitative superdiffusive transport and atomistic spin-model calculations to describe a spin-valve-type trilayer consisting of Fe$|$Cu$|$Mn$_2$Au. Our results demonstrate that femtosecond spin-transfer torques can switch the Mn$_2$Au layer within a few picoseconds. In addition,  we find that spin waves with high frequencies up to several THz can be excited in Mn$_2$Au.

\end{abstract}

\maketitle

\begin{acronym}
\acro{AFM}[AFM]{antiferromagnetic}
\acro{FM}[FM]{ferromagnetic}
\acro{NM}[NM]{nonmagnetic}
\acro{AFMR}[AFMR]{antiferromagnetic resonance}
\acro{LLG}[LLG]{Landau-Lifshitz-Gilbert}
\acro{STT}[STT]{spin-transfer torque}
\acro{SSW}[SSW]{standing spin wave}
\acro{FWHM}[FWHM]{full width at half maximum}
\end{acronym}

Antiferromagnets (AFMs)\acused{AFM} are promising materials for future spintronic devices. Among the advantages over ferromagnets (FMs)\acused{FM} are the faster spin dynamics, the lack of stray fields, the low susceptibility to magnetic fields and the abundance of materials \cite{Jungwirth2016,Baltz2018,Zelezny2018}. A challenging aspect in the field of \ac{AFM} spintronics has been for decades the fact that their order parameter is difficult to read and control, due to their lack of macroscopic magnetization. Recently, progress was made by the discovery that electrically induced N{\'e}el-spin-orbit torques \cite{Zelezny2014} can be used to switch the magnetic order in a certain class of \acp{AFM} with broken inversion symmetry, such as CuMnAs \cite{Wadley2016,Olejnik2017,Olejnik2018} and Mn$_2$Au \cite{Zelezny2014,Roy2016,Bodnar2018,Meinert2018,Salemi2019,Selzer2022}. 
{An important issue for future device applications is to know in what way, and how fast, such switching process could best proceed.}

A different line of research showed recently that femtosecond laser excitation  of a \ac{FM} creates a burst of  spin-polarized current that contributes substantially and in a nonlocal fashion to its ultrafast demagnetization \cite{Battiato2010,Malinowski2008,Melnikov2011,Rudolf2012,Eschenlohr2013,Vodungbo2016,Bergeard2016,Xu2017,Alekhin2017}. Moreover, it was demonstrated that these spin-current pulses can give rise to an ultrafast \ac{STT} \cite{Schellekens2014,Choi2014} and excite high-frequency spin waves in an adjacent Fe layer \cite{Razdolski2017,Ritzmann2020}. 
Earlier works have suggested that \acp{STT} arising from spin currents transmitted through \ac{AFM} layers can induce large torques and even switching \cite{Gomonay2010,Chirac2020}. {An intriguing question is what happens when  ultrashort spin-current pulses act on an AFM.}

\begin{figure}
    \centering
    \includegraphics[width=0.47\textwidth]{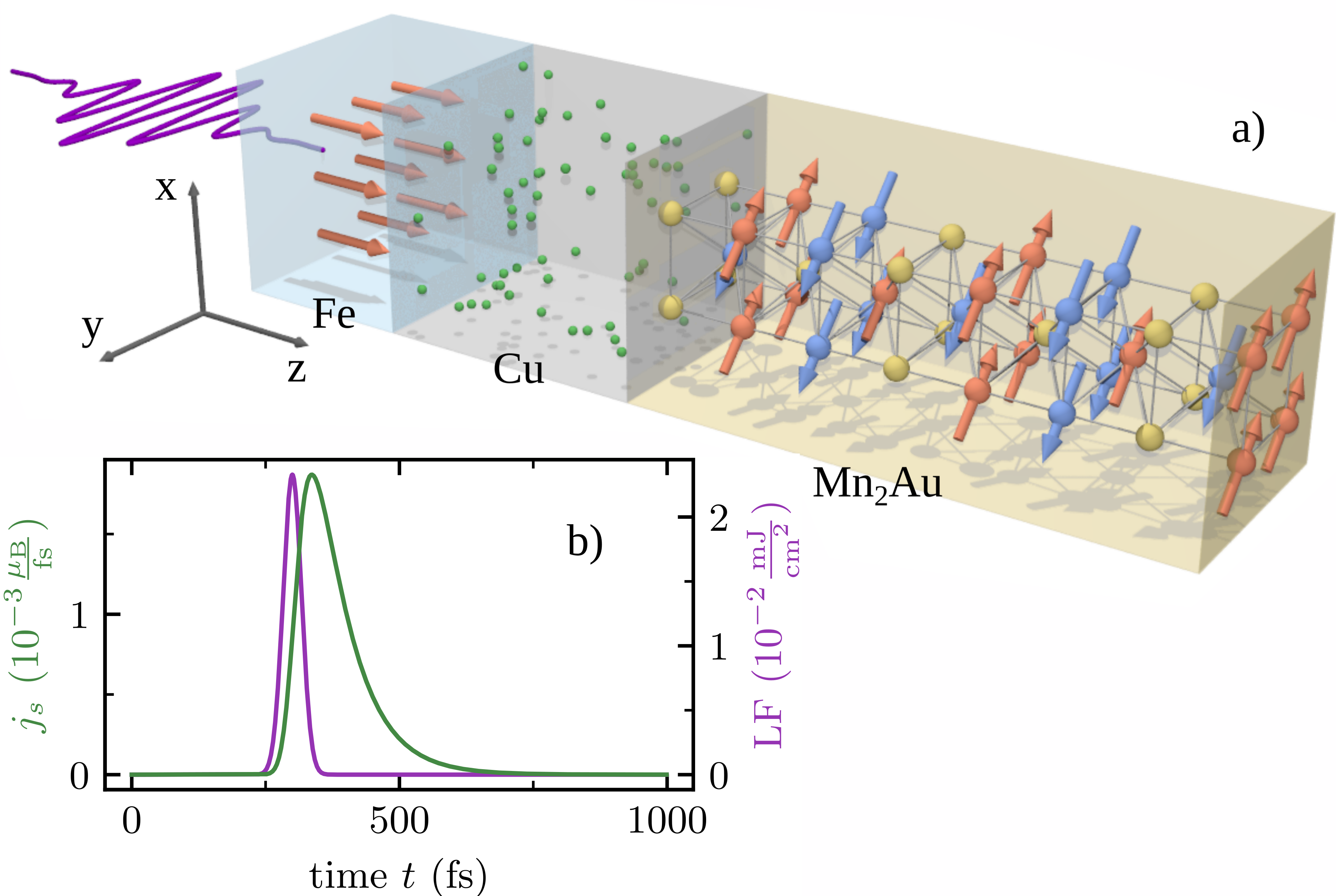}
    \caption{(a) Sketch of the studied trilayer structure Fe$|$Cu$|$Mn$_2$Au. Magnetic moments are represented by arrows. The Fe layer is excited by an ultrafast laser pulse (purple). The demagnetization of Fe generates a spin current of hot electrons (green balls), which is transmitted through the Cu layer into Mn$_2$Au, where it exerts a femtosecond \ac{STT} on the magnetic moments. The two Mn sublattices are illustrated by red and blue spheres and the Au atoms are shown in gold. (b) Laser fluence (LF) of the laser pulse with total fluence of $\SI{1}{\milli\joule/\cm\squared}$ and a \acs*{FWHM} of $\SI{40}{\femto\second}$ and calculated superdiffusive spin-current $j_\mathrm{s}$ per atom at the interface between Cu and Mn$_2$Au. Note that the maximum of the laser pulse is at $t_0 = \SI{300}{\femto\second}$.}
    \label{fig:setup}
\end{figure}

Here, we present a quantitative theoretical study to investigate \ac{AFM} dynamics due to femtosecond \acp{STT} emerging from laser-induced demagnetization of a \ac{FM}. We combine superdiffusive spin-transport calculations and {\textit{ab initio} parametrized \cite{Selzer2022}}  atomistic spin-dynamics simulations to study a spin-valve-type trilayer consisting of Fe$|$Cu$|$Mn$_2$Au, see Fig.~\ref{fig:setup}(a). Solving numerically the atomistic \ac{LLG} \cite{Nowak2007} we demonstrate the formation of thickness-dependent spin-wave spectra with significant peaks at frequencies of up to several $\si{\tera\hertz}$.
{Remarkably,} we reveal that the laser-induced spin currents can induce {N\'eel-vector} switching {within $\SI{2}{\pico\second}$}, opening up a new pathway for the ultrafast and efficient control of the magnetic order in \acp{AFM}.

\textit{Methodology.}
To model the spin current emerging due to the ultrafast laser-induced demagnetization of Fe, we use the superdiffusive spin-transport theory \cite{Battiato2010} and its extension to magnetic heterostructures consisting of a \ac{FM} and a \ac{NM} layer \cite{Battiato2012}. The model assumes that the laser pulse excites two channels of spin up and down electrons, respectively, in the \ac{FM} layer. Since in the \ac{FM} layer the two channels have different transport properties (i.e.,\ lifetime and velocities of electrons), the resulting net current is a spin-polarized current which {strongly} 
{contributes to} the femtosecond demagnetization of the FM layer when the current is injected in the \ac{NM} layer, {see Supplementary Material (SM) for details
\cite{SupplMat}.}

We model the external laser source $S^\mathrm{ext}_\sigma$ that generates the nonthermal electron population as a Gaussian pulse $    S_\sigma(E,z)\propto N_\sigma(E,z)\mathrm{exp}{\{-\frac{(t-t_0)^2}{2\Delta^2}\}},     $ where $t_0$ is the time position of the pulse peak, $\Delta$ is the standard deviation of the pulse and $N_\sigma(E,z)$ is the number of excited electrons {per spin $\sigma$} in the material. In particular, we assume that the laser has a finite penetration length ($\lambda_\mathrm{laser}$), so that the number of excited electrons decays far from the surface $N_\sigma(E,z)=N_0\exp{(-z/\lambda_\mathrm{laser})}$, and $N_0$ is a quantity directly proportional to the laser fluence.

Solving the superdiffusive transport equation gives the spin-current density $j_\mathrm{s}(z,t)$, {shown in} Fig.~\ref{fig:setup}(b), which is defined as the difference between spin-up and spin-down electrons flowing in and out at each position $z$ \cite{Lu2020}. 

All results presented below were obtained for laser pulses with \ac{FWHM} \footnote{The FWHM of a Gaussian pulse is related to its standard deviation via $\mathrm{FWHM}\approx 2.355 \Delta$} of $\SI{40}{\femto\second}$ and thicknesses of $\SI{16}{\nano\metre}$ and $\SI{4}{\nano\metre}$ of the Fe and Cu films, respectively. The impact of varying the \ac{FWHM} on the spin-current pulse and the switching dynamics is discussed in the 
SM~\cite{SupplMat}.

When the current $j_\mathrm{s}$ is transmitted from the Cu layer into Mn$_2$Au it exerts a \ac{STT} \cite{Slonzweski1996,Berger1996,Slonzweski2002} on the Mn moments. In order to describe the time evolution of the localized Mn moments under this torque, we numerically solve the \ac{LLG} equation \cite{Nowak2007,Kazantseva2008},
\begin{align}
    \begin{split}
    \frac{\partial \vec{S}_i}{\partial t}
    =
    &-
    \frac{\gamma}{\mu_\mathrm{s}}
    \vec{S}_i
    \times
    \vec{H}^\mathrm{eff}_i
    +
    \alpha
    \vec{S}_i
    \times
    \frac{\partial \vec{S}_i}{\partial t}
    \\
    &+
    \frac{j_\mathrm{s}(z,t)}{\mu_\mathrm{s}}
    \vec{S}_i
    \times
    (
    \vec{S}_i
    \times
    \hat{
    \vec{z}
    }
    ).
    \end{split}
    \label{eq:LLG}
\end{align}
$\vec{S}_i$
are the normalized magnetic moments, $\mu_\mathrm{s}=\num{3.74}\,\mu_\mathrm{B}$ is the saturation magnetic moment \cite{Selzer2022}, 
$\gamma = \SI{1.76e11}{\second^{-1}\tesla^{-1}}$ is the absolute value of the gyromagnetic ratio and $\alpha$ is the dimensionless Gilbert damping parameter. The effective field $\vec{H}^\mathrm{eff}_i= -\partial\mathcal{H}/\partial \vec{S}_i$ is generated by exchange interaction with neighboring Mn moments and anisotropy terms (see below). 

We also consider an additional {(3$^{\rm rd}$)} term due to the femtosecond \ac{STT}, which acts as an antidamping torque term and is a result of the absorption of the spin current in Mn$_2$Au \cite{Gomonay2010,Balaz2018,Chirac2020}. This term is proportional to the spin-current amplitude $j_\mathrm{s}$ of the hot electrons at the interface between Cu and Mn$_2$Au, the temporal evolution of which we calculate using the superdiffusive spin-transport theory described above. Note that the polarization of the spin current is along the $\hat{\vec{z}}$ direction. The typical length scale at which the spin current is absorbed is at most a few nanometers \cite{Stiles2002,Ghosh2012,Razdolski2017}. Following Ref.~\cite{Ritzmann2020}, we assume spin-current absorption within a characteristic penetration depth $\lambda_\mathrm{STT}$,
\begin{align}
    j_\mathrm{s}(z,t)
    =
    j_\mathrm{s}(0,t)
    \frac{\exp(-z/\lambda_\mathrm{STT})}{\sum_z \exp(-z/\lambda_\mathrm{STT})}
    \label{eq:spin_current_spatial}.
\end{align}
As a general feature of our simulations, we found that larger penetration depths lead to less pronounced peaks in the frequency spectra, as was also demonstrated in Ref.~\cite{Ritzmann2020}, and that they obstruct efficient switching (the latter aspect is also discussed below). Unless stated otherwise, all results presented hereinafter are for $\lambda_\mathrm{STT}=\SI{1}{\nano\meter}$, 
comparable to the value {used} for Fe 
\cite{Ritzmann2020}.

The spin-model Hamiltonian for Mn$_2$Au we use here has been parameterized in Ref.~\cite{Selzer2022} using \textit{ab initio} calculations and reads

\begin{align}
\begin{split}
\mathcal{H}
=
&-
\frac{1}{2}
\sum_{i\neq j}
J_{ij}
\vec{S}_i
\cdot
\vec{S}_j
-
d_z
\sum_{i}
 {S}_{i,z}^2\\
&-
d_{zz}
\sum_{i}
{S}_{i,z}^4
-
d_{xy}
\sum_{i}
{S}_{i,x}^2
{S}_{i,y}^2.
\end{split}
\label{eq:Hamiltonian}
\end{align}
This Hamiltonian includes exchange between Mn moments at different lattice site $i$ and $j$ beyond nearest neighbors and the anisotropy terms reflect the tetragonal symmetry of the unit cell. The exchange parameters $J_{ij}$ alternate in sign depending on distance between the two magnetic moments 
\cite{Selzer2022}. The values of the anisotropy constants are $d_{z}= -\SI{0.62}{\milli\electronvolt}$, $d_{zz}= -\SI{0.024}{\milli\electronvolt}$ and $d_{xy}= \SI{0.058}{\milli\electronvolt}$. These parameters give rise to a layered \ac{AFM} groundstate where the N{\'e}el vectors $\vec{n}_i = (\vec{m}_{2i-1} -\vec{m}_{2i})/2$, with $\vec{m}_\mathrm{2i-1/2i}$ being the alternating magnetization (i.e., the sum over the spins in one layer) of the layers along the $z$ axis, are aligned collinearly along the diagonals of the $x$-$y$-plane (shown in Fig.~\ref{fig:setup}). Given the symmetry of Mn$_2$Au, the states with the N{\'e}el vector along the crystallographic directions $[110]$, $[\bar{1}10]$, $[1\bar{1}0]$ and $[\bar{1}\bar{1}0]$ are equivalent. Note that {solving} the model used here \cite{note} predicts a critical temperature of $\SI{1680}{\kelvin}$ \cite{Selzer2022}, which is in reasonable agreement with what was reported experimentally \cite{Barthem2013}.

\begin{figure}[t]
    \centering
    \includegraphics[width=0.45\textwidth]{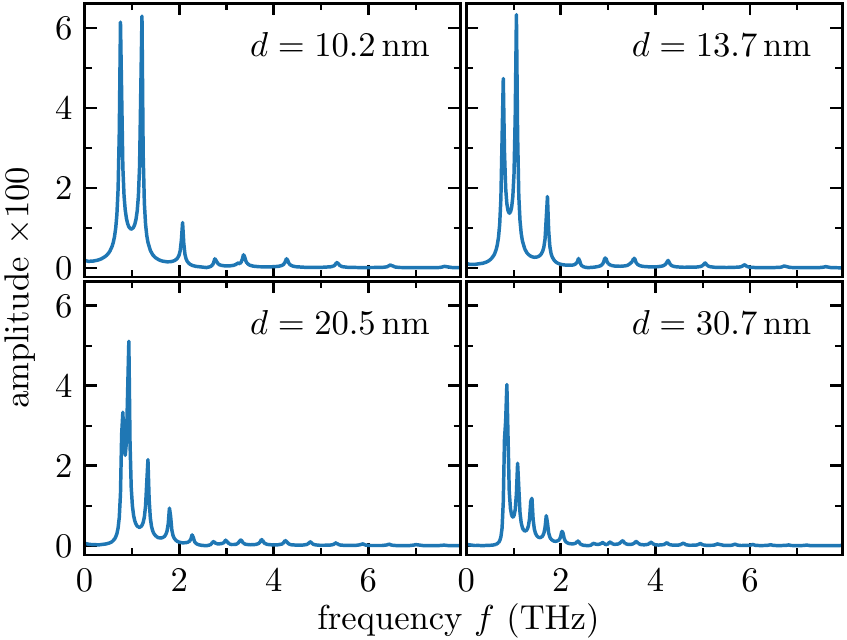}
    \caption{Amplitudes of excited spin waves (see text) as a function of frequency for different Mn$_2$Au thicknesses.}
    \label{fig:spectra}
\end{figure}

\textit{THz spin-wave excitation.}
To begin with, we investigate the formation of \acp{SSW} that are excited by a laser pulse with a fluence of $\SI{0.54}{\milli\joule/\cm\squared}$ during the first $\SI{40}{\pico\second}$, {for Mn$_2$Au layers with thicknesses ranging from $\SI{853.9}{\pico\metre}$ to $\SI{30.7}{\nano\metre}$}. The femtosecond \ac{STT} excites high-frequency spin waves, which propagate through the Mn$_2$Au layer and can be reflected multiple times before decaying. 
We obtain the spectrum of appearing frequencies by performing a Fourier transformation on the time domain of the magnetization in the last layer (note that all spins in the last layer belong to the same sublattice). In Fig.~\ref{fig:spectra} we show the spin-wave frequency spectra for different thicknesses of Mn$_2$Au. To obtain the amplitude, we calculate the 
vector length consisting of the absolute values of Fourier transforms of the magnetization in the last layer, i.e.\ $ [|\tilde{m}_x(f)|^2 + |\tilde{m}_y(f)|^2 +|\tilde{m}_z(f)|^2 ]^{1/2}$ with 
\begin{align}
    \tilde{\vec{m}}(f) 
    =
    \frac{1}{\sqrt{N_\mathrm{steps}}}
    \sum_{n=0}^{N_\mathrm{steps}-1} \vec{m}_N(t_n)\exp(-\mathrm{i}2\pi f t_n),
\end{align}
where $N_\mathrm{steps}$ is the number of time steps and $t_n = n\Delta t$. The frequency spectra reveal multiple peaks indicating the formation of \acp{SSW} of up to several $\si{\tera\hertz}$. 
{Such high-frequency spin-wave modes have become attractive recently for spintronics operating at THz frequencies  \cite{Vaidya2020,Li2020,Salikhov2023}.}
The lowest lying peak represents an \ac{AFMR} mode and appears for all shown thicknesses at the same frequency of about $\SI{0.8}{\tera\hertz}$. By increasing the thickness $d$ of the Mn$_2$Au layer, the number of peaks in the displayed range increases since the interval between the peaks decreases (for $d=\SI{30.7}{\nano\metre}$, the first two peaks even become indistinguishable). This is consistent with the fact that, as a general aspect of standing wave formation, the 
{thickness $d$} of the propagation medium is a multiple of the allowed wavelengths and, henceforth, that the allowed wave vectors scale inversely, 
i.e.~$k_n=n \pi/d$. Since dispersion relations typically feature a monotonic increase in frequency with absolute value of the wave vector, this results in an increased number of peaks within a certain interval for thicker Mn$_2$Au layers. Note that 
it is {impossible to construct the dispersion relation} for spin waves in Mn$_2$Au solely from the frequency spectra shown in Fig.~\ref{fig:spectra} and the condition for \ac{SSW}, $k_n=n \pi/d$. This is because the dispersion relation of Mn$_2$Au has more than one branch, which obstructs the unambiguous identification of the $n$-th peak in the spectrum as belonging to the wave vector $k_n$ (we discuss this in more detail in the SM~\cite{SupplMat}).

{Instead, the dispersion relation} of Mn$_2$Au can be obtained using linear spin-wave theory. In the vicinity of the groundstate, the Hamiltonian~\eqref{eq:Hamiltonian} can be mapped onto a biaxial system with the easy axis along the $[110]$ direction, with an easy axis anisotropy value of $d_{xy}$, while the hard axis keeps its orientation along the $z$ axis. If only (\ac{AFM}) nearest-neighbor exchange is assumed, we can use {literature} formulas 
for the two emerging \ac{AFMR} modes \cite{Rezende2019},
\begin{align}
    f_0^\mathrm{a} &= \frac{\gamma}{2 \pi\mu_\mathrm{s}} 
    [ 2J^\mathrm{inter}d_{xy}]^{1/2} \approx \SI{0.85}{\tera\hertz} ,\\
    f_0^\mathrm{b} &= \frac{\gamma}{2 \pi\mu_\mathrm{s}} [ 2J^\mathrm{inter}(d_{xy}-d_z) ]^{1/2} \approx \SI{2.9}{\tera\hertz}.
\end{align}
The effective inter-sublattice exchange coupling {obtained by summing up interactions up to a spatial cutoff of $\SI{0.9}{\nano\metre}$} is given by $J^\mathrm{inter} = \SI{371.13}{\milli\electronvolt}$ \cite{Selzer2022}. Albeit the approximations described above, the value for $f_0^\mathrm{a}$ agrees well with the simulation results. 
Based on our calculations within linear spin-wave theory, we would expect a second thickness-independent \ac{AFMR} peak at $f_0^\mathrm{b}$ in the spectra. However, due to the finite linewidth and the density of the peaks in the vicinity of $f_0^\mathrm{b}$, such a peak cannot unambiguously be identified.

\begin{figure}[t!]
    \centering
    \includegraphics[width=0.45\textwidth]{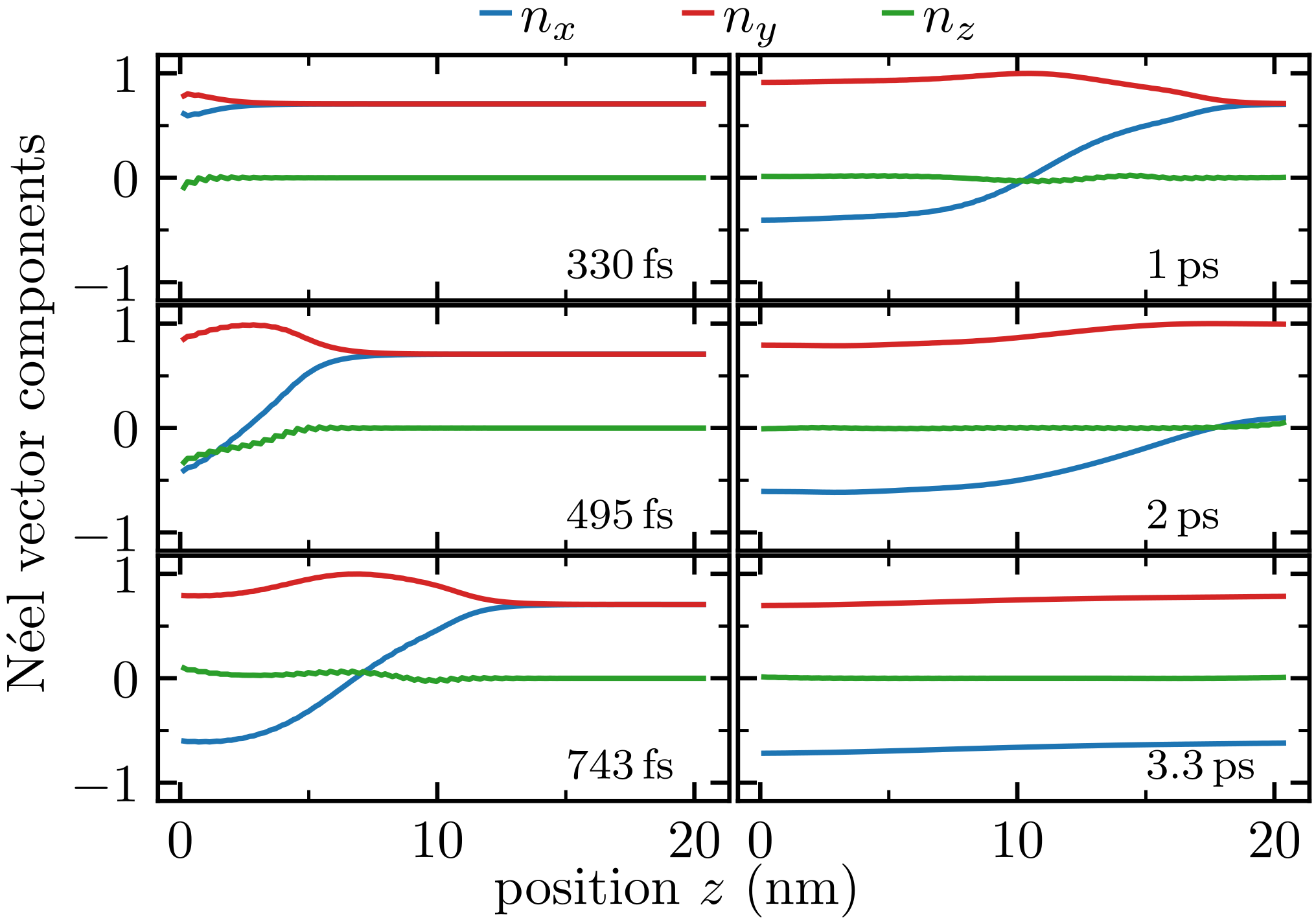}
    \caption{{Components of the N{\'e}el vector $\vec{n}_i$} of Mn$_2$Au versus distance from the interface to Cu (located at $z=0$)  during the switching process from $[110]$ to $[\bar{1}10]$ at different points in time. The laser pulse starts at $t=0$ and the \ac{STT} starts at about $\SI{300}{\femto\second}$.
    The laser fluence has a value of $\SI{14.86}{\milli\joule/\cm\squared}$ and the damping used is $\alpha=0.01$.}
    \label{fig:switching_cs}
\end{figure}

\textit{AFM switching.}
Upon increasing the laser fluence, the femtosecond \ac{STT} due to the emerging spin current eventually becomes strong enough to drive the magnetic moments near the interface over the energy barrier induced by the in-plane anisotropy $d_{xy}$. This excitation propagates through the Mn$_2$Au layer and can ultimately lead to $90^{\circ}$ switching of the N{\'e}el vector at the timescale of few picoseconds, see Fig.~\ref{fig:switching_cs}.

This rapid switching is a result of the so-called exchange enhancement, which is characteristic for \ac{AFM} dynamics \cite{Kittel1951,Gomonay2014,Roy2016,Dannegger2021,Selzer2022}. The \ac{STT} in the \ac{LLG} {Eq.\ \eqref{eq:LLG}}
is quadratic in the magnetization and thus has the same direction on each sublattice. This gives rise to a canting between the sublattices and the emerging inter-sublattice exchange field leads {then} to fast precessional motion of the magnetic moments.

\begin{figure}[t]
    \centering
    \includegraphics[width=0.45\textwidth]{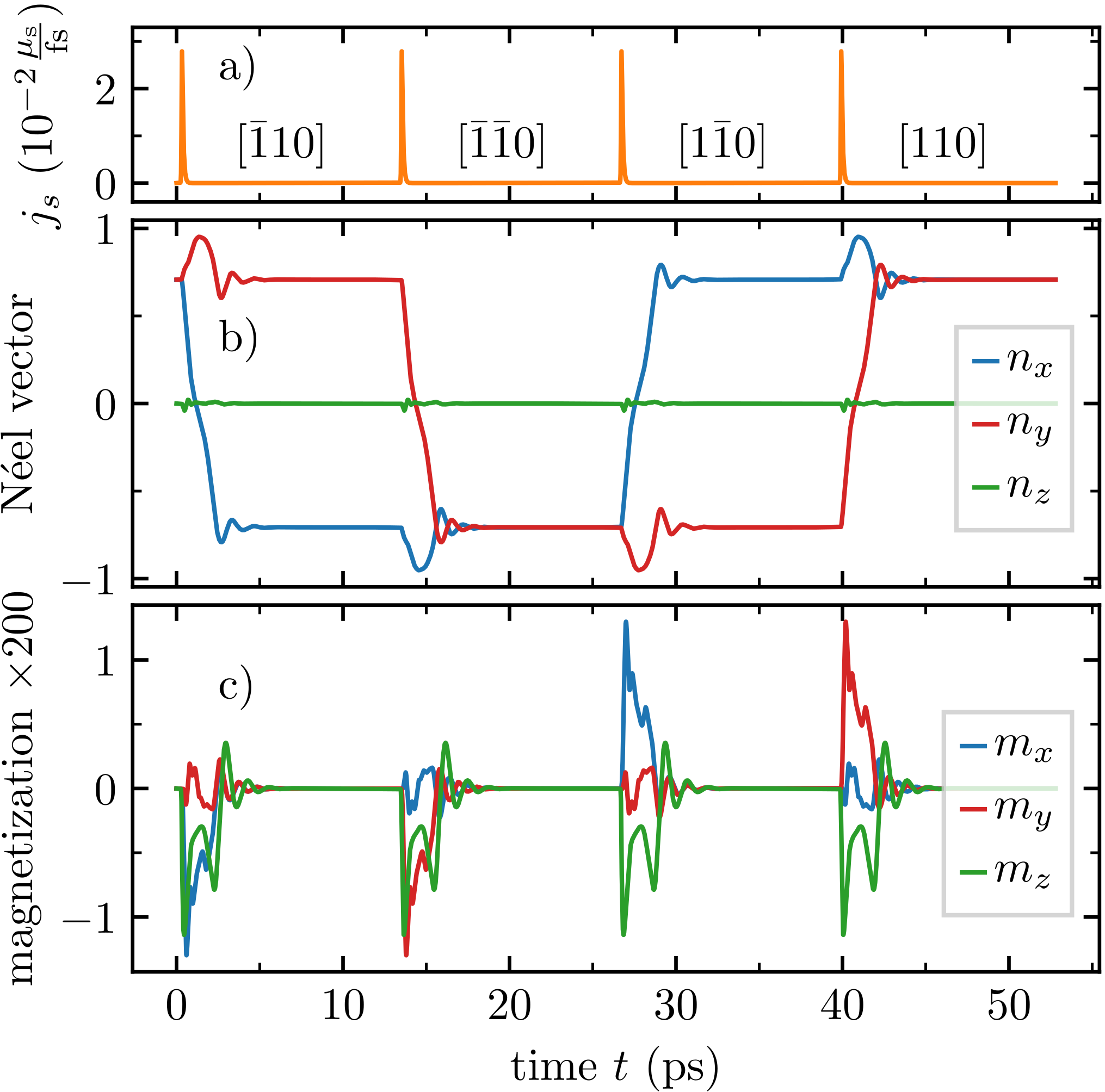}
    \caption{Cyclic switching of Mn$_2$Au. (a) Spin current resulting from a sequence of four laser pulses with a laser fluence of $\SI{14.86}{\milli\joule/\cm\squared}$ every $\SI{13.2}{\pico\second}$. (b) Components of {the} average N{\'e}el vector $\vec{n}=(N/2)^{-1}\sum_{i=1}^{N/2} \vec{n}_i$,
     with $N$ being the number of Mn layers. (c) Components of average magnetization $\vec{m}=(N)^{-1}\sum_{i=1}^{N}\vec{m}_i$. The system has a thickness of $\SI{20.5}{\nano\metre}$ and the Gilbert damping parameter is set to $\alpha=0.01$.}
    \label{fig:cyclic_switching}
\end{figure}

{Next, we show in Fig.~\ref{fig:cyclic_switching} that} repeated laser pulses at the same fluence lead to cyclic switching of the Mn$_2$Au layer. 
Each excitation switches the N{\'e}el {vector}
by $90^{\circ}$ in counter-clockwise direction around the $z$ axis. Note that reversing the polarization of the Fe layer leads to a sign change in the femtosecond \ac{STT} in Eq.~\eqref{eq:LLG} and, hence, to switching in clockwise direction. The sublattice canting is visible in a small, but finite total magnetization {during switching}.

\begin{figure}
    \centering
    \includegraphics[width=0.47\textwidth]{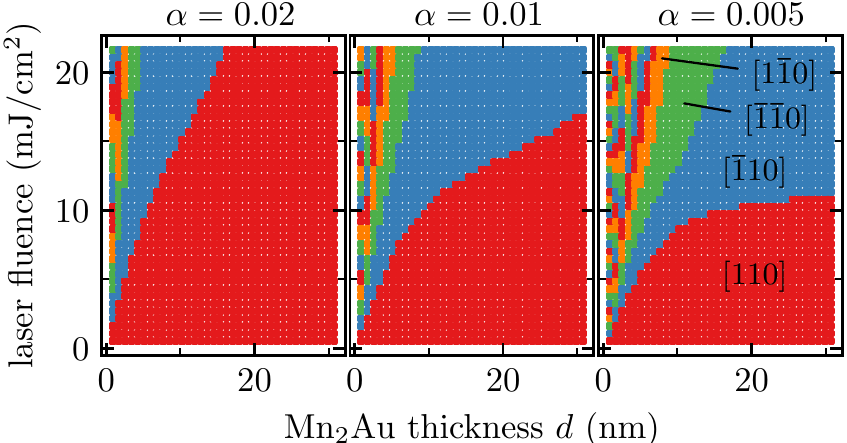}
    \caption{Switching {phase diagram} of Mn$_2$Au due to femtosecond \acp{STT} for varying laser fluences, thicknesses and values of the Gilbert damping parameter $\alpha$ as labeled. The color coding describes the orientation of the average N{\'e}el vector after the laser excitation. Initially, the N{\'e}el vector is {along} the $[110]$ direction.}
    \label{fig:switching_vs_LF_and_d}
\end{figure}

Further increase of the laser fluence leads to more canting, so that the precessional motion of the magnetic moments close to the interface persists for a longer time. Depending on the timescale of the relaxation back to antiparallel alignment of the sublattice magnetizations - which crucially depends on the Gilbert damping parameter $\alpha$ - the rotation of the N{\'e}el order parameter can even be larger than $90^{\circ}$. On the other hand, a larger thickness $d$ of the Mn$_2$Au layer can obstruct the switching because the energy barrier that needs to be overcome scales linearly with $d$, whereas the total torque due to the absorption of the spin current close to the interface does not depend on $d$, as long as $d$ is not comparable to the penetration depth. These features of the switching dynamics are summarized in Fig.~\ref{fig:switching_vs_LF_and_d}.

We also want to point out the key role of the highly localized absorption of the spin current in the vicinity of the Cu/Mn$_2$Au interface. Our results indicate that in general larger penetration depths hinder efficient switching, with greatly decreased dynamics in the limit of $\lambda_\mathrm{STT}\rightarrow \infty$, i.e.\ for spatially homogeneous \ac{STT}, and much larger fluences are required for switching. As such, describing the switching of \ac{AFM} layers by ultrafast \ac{STT} using a two-sublattice macrospin model, as done in Ref.~\cite{Chirac2020} for NiO, greatly overestimates the threshold laser fluences as compared to the spatially resolved spin dynamics simulations done here.
In addition, our findings indicate that the timescales at which the \acp{STT} are present have a decisive impact on the switching threshold: for a constant total laser fluence, higher \acp{FWHM} obstruct switching. These aspects are analyzed in more detail in the SM~\cite{SupplMat}.

In Ref.~\cite{Selzer2022} it was demonstrated that thermal activation plays a crucial role in the switching of Mn$_2$Au via N{\'e}el-spin-orbit torques, leading to -- in some cases -- zero 
switching {probability} in the absence of thermal fluctuations and almost deterministic switching at temperatures slightly above room temperature.
Here we find, using the stochastic \ac{LLG} \cite{Nowak2007}, that the {temperature} enhancement of switching probability 
is less pronounced; e.g., for $d=\SI{20.5}{\nano\metre}$, $\alpha=0.01$ and a laser fluence of $\SI{12.97}{\milli\joule/\cm\squared}$ (which is slightly below the threshold fluence for switching, see Fig.~\ref{fig:switching_vs_LF_and_d}), the switching probability only increases up to around $15\%$ at room temperature and to about $35\%$ at $\SI{600}{\kelvin}$. This {difference} can be attributed to the fact that here the torques act on the timescale of femtoseconds, in contrast to the $\SI{20}{\pico\second}$ in Ref.~\cite{Selzer2022}. The impact of temperature on the switching dynamics is discussed in more detail in the SM~\cite{SupplMat}.

To summarize, we have studied theoretically the laser-induced excitation of $\si{\tera\hertz}$ spin waves and switching in an Fe$|$Cu$|$Mn$_2$Au trilayer structure. The spin current emerging from the laser-induced ultrafast demagnetization of Fe was calculated using the superdiffusive spin-transport theory. These spin-current pulses excite spin dynamics in \ac{AFM} Mn$_2$Au via femtosecond \acp{STT} that were modeled using \textit{ab initio} parametrized atomistic spin-dynamics simulations. Our results reveal the formation of thickness-dependent  frequency spectra, demonstrating the formation of \acp{SSW} with frequencies of up to several $\si{\tera\hertz}$. At larger laser fluences, the spin-current pulse leads to ultrafast  switching of the Mn$_2$Au layer.
Our quantitative findings predict an efficient way to excite \acp{AFM} dynamics -- $\si{\tera\hertz}$ spin waves and switching -- using trilayer structures and femtosecond laser pulses. As such, they open up a new pathway for the efficient and ultrafast manipulation of magnetic order in antiferromagnets.

{We thank Tobias Dannegger and L{\'a}szl{\'o} Szunyogh for helpful discussions. This work was supported by the DFG (Deutsche Forschungsgemeinschaft) via TRR 227 ``Ultrafast Spin Dynamics" (Project MF), the Swedish Research Council (VR) (Grants No.\ 2018-05973 and 2021-05211),  the K.\ and A.\ Wallenberg Foundation (Grant No.\ 2022.0079), and the European Union’s Horizon 2020 Research and Innovation Programme under FET-OPEN Grant Agreement No.\ 863155 (s-Nebula).
}



\begin{thebibliography}{48}%
\makeatletter
\providecommand \@ifxundefined [1]{%
 \@ifx{#1\undefined}
}%
\providecommand \@ifnum [1]{%
 \ifnum #1\expandafter \@firstoftwo
 \else \expandafter \@secondoftwo
 \fi
}%
\providecommand \@ifx [1]{%
 \ifx #1\expandafter \@firstoftwo
 \else \expandafter \@secondoftwo
 \fi
}%
\providecommand \natexlab [1]{#1}%
\providecommand \enquote  [1]{``#1''}%
\providecommand \bibnamefont  [1]{#1}%
\providecommand \bibfnamefont [1]{#1}%
\providecommand \citenamefont [1]{#1}%
\providecommand \href@noop [0]{\@secondoftwo}%
\providecommand \href [0]{\begingroup \@sanitize@url \@href}%
\providecommand \@href[1]{\@@startlink{#1}\@@href}%
\providecommand \@@href[1]{\endgroup#1\@@endlink}%
\providecommand \@sanitize@url [0]{\catcode `\\12\catcode `\$12\catcode
  `\&12\catcode `\#12\catcode `\^12\catcode `\_12\catcode `\%12\relax}%
\providecommand \@@startlink[1]{}%
\providecommand \@@endlink[0]{}%
\providecommand \url  [0]{\begingroup\@sanitize@url \@url }%
\providecommand \@url [1]{\endgroup\@href {#1}{\urlprefix }}%
\providecommand \urlprefix  [0]{URL }%
\providecommand \Eprint [0]{\href }%
\providecommand \doibase [0]{http://dx.doi.org/}%
\providecommand \selectlanguage [0]{\@gobble}%
\providecommand \bibinfo  [0]{\@secondoftwo}%
\providecommand \bibfield  [0]{\@secondoftwo}%
\providecommand \translation [1]{[#1]}%
\providecommand \BibitemOpen [0]{}%
\providecommand \bibitemStop [0]{}%
\providecommand \bibitemNoStop [0]{.\EOS\space}%
\providecommand \EOS [0]{\spacefactor3000\relax}%
\providecommand \BibitemShut  [1]{\csname bibitem#1\endcsname}%
\let\auto@bib@innerbib\@empty
\bibitem [{\citenamefont {Jungwirth}\ \emph {et~al.}(2016)\citenamefont
  {Jungwirth}, \citenamefont {Marti}, \citenamefont {Wadley},\ and\
  \citenamefont {Wunderlich}}]{Jungwirth2016}%
  \BibitemOpen
  \bibfield  {author} {\bibinfo {author} {\bibfnamefont {T.}~\bibnamefont
  {Jungwirth}}, \bibinfo {author} {\bibfnamefont {X.}~\bibnamefont {Marti}},
  \bibinfo {author} {\bibfnamefont {P.}~\bibnamefont {Wadley}}, \ and\ \bibinfo
  {author} {\bibfnamefont {J.}~\bibnamefont {Wunderlich}},\ }\href {\doibase
  10.1038/nnano.2016.18} {\bibfield  {journal} {\bibinfo  {journal} {Nature
  Nanotechnology}\ }\textbf {\bibinfo {volume} {11}},\ \bibinfo {pages} {231}
  (\bibinfo {year} {2016})}\BibitemShut {NoStop}%
\bibitem [{\citenamefont {Baltz}\ \emph {et~al.}(2018)\citenamefont {Baltz},
  \citenamefont {Manchon}, \citenamefont {Tsoi}, \citenamefont {Moriyama},
  \citenamefont {Ono},\ and\ \citenamefont {Tserkovnyak}}]{Baltz2018}%
  \BibitemOpen
  \bibfield  {author} {\bibinfo {author} {\bibfnamefont {V.}~\bibnamefont
  {Baltz}}, \bibinfo {author} {\bibfnamefont {A.}~\bibnamefont {Manchon}},
  \bibinfo {author} {\bibfnamefont {M.}~\bibnamefont {Tsoi}}, \bibinfo {author}
  {\bibfnamefont {T.}~\bibnamefont {Moriyama}}, \bibinfo {author}
  {\bibfnamefont {T.}~\bibnamefont {Ono}}, \ and\ \bibinfo {author}
  {\bibfnamefont {Y.}~\bibnamefont {Tserkovnyak}},\ }\href {\doibase
  10.1103/RevModPhys.90.015005} {\bibfield  {journal} {\bibinfo  {journal}
  {Rev. Mod. Phys.}\ }\textbf {\bibinfo {volume} {90}},\ \bibinfo {pages}
  {015005} (\bibinfo {year} {2018})}\BibitemShut {NoStop}%
\bibitem [{\citenamefont {{\v{Z}}elezn{\'y}}\ \emph {et~al.}(2018)\citenamefont
  {{\v{Z}}elezn{\'y}}, \citenamefont {Wadley}, \citenamefont {Olejn{\'i}k},
  \citenamefont {Hoffmann},\ and\ \citenamefont {Ohno}}]{Zelezny2018}%
  \BibitemOpen
  \bibfield  {author} {\bibinfo {author} {\bibfnamefont {J.}~\bibnamefont
  {{\v{Z}}elezn{\'y}}}, \bibinfo {author} {\bibfnamefont {P.}~\bibnamefont
  {Wadley}}, \bibinfo {author} {\bibfnamefont {K.}~\bibnamefont {Olejn{\'i}k}},
  \bibinfo {author} {\bibfnamefont {A.}~\bibnamefont {Hoffmann}}, \ and\
  \bibinfo {author} {\bibfnamefont {H.}~\bibnamefont {Ohno}},\ }\href {\doibase
  10.1038/s41567-018-0062-7} {\bibfield  {journal} {\bibinfo  {journal} {Nature
  Physics}\ }\textbf {\bibinfo {volume} {14}},\ \bibinfo {pages} {220}
  (\bibinfo {year} {2018})}\BibitemShut {NoStop}%
\bibitem [{\citenamefont {\ifmmode~\check{Z}\else \v{Z}\fi{}elezn\'y}\ \emph
  {et~al.}(2014)\citenamefont {\ifmmode~\check{Z}\else \v{Z}\fi{}elezn\'y},
  \citenamefont {Gao}, \citenamefont {V\'yborn\'y}, \citenamefont {Zemen},
  \citenamefont {Ma\ifmmode~\check{s}\else \v{s}\fi{}ek}, \citenamefont
  {Manchon}, \citenamefont {Wunderlich}, \citenamefont {Sinova},\ and\
  \citenamefont {Jungwirth}}]{Zelezny2014}%
  \BibitemOpen
  \bibfield  {author} {\bibinfo {author} {\bibfnamefont {J.}~\bibnamefont
  {\ifmmode~\check{Z}\else \v{Z}\fi{}elezn\'y}}, \bibinfo {author}
  {\bibfnamefont {H.}~\bibnamefont {Gao}}, \bibinfo {author} {\bibfnamefont
  {K.}~\bibnamefont {V\'yborn\'y}}, \bibinfo {author} {\bibfnamefont
  {J.}~\bibnamefont {Zemen}}, \bibinfo {author} {\bibfnamefont
  {J.}~\bibnamefont {Ma\ifmmode~\check{s}\else \v{s}\fi{}ek}}, \bibinfo
  {author} {\bibfnamefont {A.}~\bibnamefont {Manchon}}, \bibinfo {author}
  {\bibfnamefont {J.}~\bibnamefont {Wunderlich}}, \bibinfo {author}
  {\bibfnamefont {J.}~\bibnamefont {Sinova}}, \ and\ \bibinfo {author}
  {\bibfnamefont {T.}~\bibnamefont {Jungwirth}},\ }\href {\doibase
  10.1103/PhysRevLett.113.157201} {\bibfield  {journal} {\bibinfo  {journal}
  {Phys. Rev. Lett.}\ }\textbf {\bibinfo {volume} {113}},\ \bibinfo {pages}
  {157201} (\bibinfo {year} {2014})}\BibitemShut {NoStop}%
\bibitem [{\citenamefont {Wadley}\ \emph {et~al.}(2016)\citenamefont {Wadley},
  \citenamefont {Howells}, \citenamefont {Železný}, \citenamefont {Andrews},
  \citenamefont {Hills}, \citenamefont {Campion}, \citenamefont {Novák},
  \citenamefont {Olejník}, \citenamefont {Maccherozzi}, \citenamefont {Dhesi},
  \citenamefont {Martin}, \citenamefont {Wagner}, \citenamefont {Wunderlich},
  \citenamefont {Freimuth}, \citenamefont {Mokrousov}, \citenamefont {Kuneš},
  \citenamefont {Chauhan}, \citenamefont {Grzybowski}, \citenamefont
  {Rushforth}, \citenamefont {Edmonds}, \citenamefont {Gallagher},\ and\
  \citenamefont {Jungwirth}}]{Wadley2016}%
  \BibitemOpen
  \bibfield  {author} {\bibinfo {author} {\bibfnamefont {P.}~\bibnamefont
  {Wadley}}, \bibinfo {author} {\bibfnamefont {B.}~\bibnamefont {Howells}},
  \bibinfo {author} {\bibfnamefont {J.}~\bibnamefont {Železný}}, \bibinfo
  {author} {\bibfnamefont {C.}~\bibnamefont {Andrews}}, \bibinfo {author}
  {\bibfnamefont {V.}~\bibnamefont {Hills}}, \bibinfo {author} {\bibfnamefont
  {R.~P.}\ \bibnamefont {Campion}}, \bibinfo {author} {\bibfnamefont
  {V.}~\bibnamefont {Novák}}, \bibinfo {author} {\bibfnamefont
  {K.}~\bibnamefont {Olejník}}, \bibinfo {author} {\bibfnamefont
  {F.}~\bibnamefont {Maccherozzi}}, \bibinfo {author} {\bibfnamefont {S.~S.}\
  \bibnamefont {Dhesi}}, \bibinfo {author} {\bibfnamefont {S.~Y.}\ \bibnamefont
  {Martin}}, \bibinfo {author} {\bibfnamefont {T.}~\bibnamefont {Wagner}},
  \bibinfo {author} {\bibfnamefont {J.}~\bibnamefont {Wunderlich}}, \bibinfo
  {author} {\bibfnamefont {F.}~\bibnamefont {Freimuth}}, \bibinfo {author}
  {\bibfnamefont {Y.}~\bibnamefont {Mokrousov}}, \bibinfo {author}
  {\bibfnamefont {J.}~\bibnamefont {Kuneš}}, \bibinfo {author} {\bibfnamefont
  {J.~S.}\ \bibnamefont {Chauhan}}, \bibinfo {author} {\bibfnamefont {M.~J.}\
  \bibnamefont {Grzybowski}}, \bibinfo {author} {\bibfnamefont {A.~W.}\
  \bibnamefont {Rushforth}}, \bibinfo {author} {\bibfnamefont {K.~W.}\
  \bibnamefont {Edmonds}}, \bibinfo {author} {\bibfnamefont {B.~L.}\
  \bibnamefont {Gallagher}}, \ and\ \bibinfo {author} {\bibfnamefont
  {T.}~\bibnamefont {Jungwirth}},\ }\href {\doibase 10.1126/science.aab1031}
  {\bibfield  {journal} {\bibinfo  {journal} {Science}\ }\textbf {\bibinfo
  {volume} {351}},\ \bibinfo {pages} {587} (\bibinfo {year}
  {2016})}\BibitemShut {NoStop}%
\bibitem [{\citenamefont {Olejn{\'i}k}\ \emph {et~al.}(2017)\citenamefont
  {Olejn{\'i}k}, \citenamefont {Schuler}, \citenamefont {Marti}, \citenamefont
  {Nov{\'a}k}, \citenamefont {Ka{\v{s}}par}, \citenamefont {Wadley},
  \citenamefont {Campion}, \citenamefont {Edmonds}, \citenamefont {Gallagher},
  \citenamefont {Garces}, \citenamefont {Baumgartner}, \citenamefont
  {Gambardella},\ and\ \citenamefont {Jungwirth}}]{Olejnik2017}%
  \BibitemOpen
  \bibfield  {author} {\bibinfo {author} {\bibfnamefont {K.}~\bibnamefont
  {Olejn{\'i}k}}, \bibinfo {author} {\bibfnamefont {V.}~\bibnamefont
  {Schuler}}, \bibinfo {author} {\bibfnamefont {X.}~\bibnamefont {Marti}},
  \bibinfo {author} {\bibfnamefont {V.}~\bibnamefont {Nov{\'a}k}}, \bibinfo
  {author} {\bibfnamefont {Z.}~\bibnamefont {Ka{\v{s}}par}}, \bibinfo {author}
  {\bibfnamefont {P.}~\bibnamefont {Wadley}}, \bibinfo {author} {\bibfnamefont
  {R.~P.}\ \bibnamefont {Campion}}, \bibinfo {author} {\bibfnamefont {K.~W.}\
  \bibnamefont {Edmonds}}, \bibinfo {author} {\bibfnamefont {B.~L.}\
  \bibnamefont {Gallagher}}, \bibinfo {author} {\bibfnamefont {J.}~\bibnamefont
  {Garces}}, \bibinfo {author} {\bibfnamefont {M.}~\bibnamefont {Baumgartner}},
  \bibinfo {author} {\bibfnamefont {P.}~\bibnamefont {Gambardella}}, \ and\
  \bibinfo {author} {\bibfnamefont {T.}~\bibnamefont {Jungwirth}},\ }\href
  {\doibase 10.1038/ncomms15434} {\bibfield  {journal} {\bibinfo  {journal}
  {Nature Communications}\ }\textbf {\bibinfo {volume} {8}},\ \bibinfo {pages}
  {15434} (\bibinfo {year} {2017})}\BibitemShut {NoStop}%
\bibitem [{\citenamefont {Olejník}\ \emph {et~al.}(2018)\citenamefont
  {Olejník}, \citenamefont {Seifert}, \citenamefont {Kašpar}, \citenamefont
  {Novák}, \citenamefont {Wadley}, \citenamefont {Campion}, \citenamefont
  {Baumgartner}, \citenamefont {Gambardella}, \citenamefont {Němec},
  \citenamefont {Wunderlich}, \citenamefont {Sinova}, \citenamefont {Kužel},
  \citenamefont {Müller}, \citenamefont {Kampfrath},\ and\ \citenamefont
  {Jungwirth}}]{Olejnik2018}%
  \BibitemOpen
  \bibfield  {author} {\bibinfo {author} {\bibfnamefont {K.}~\bibnamefont
  {Olejník}}, \bibinfo {author} {\bibfnamefont {T.}~\bibnamefont {Seifert}},
  \bibinfo {author} {\bibfnamefont {Z.}~\bibnamefont {Kašpar}}, \bibinfo
  {author} {\bibfnamefont {V.}~\bibnamefont {Novák}}, \bibinfo {author}
  {\bibfnamefont {P.}~\bibnamefont {Wadley}}, \bibinfo {author} {\bibfnamefont
  {R.~P.}\ \bibnamefont {Campion}}, \bibinfo {author} {\bibfnamefont
  {M.}~\bibnamefont {Baumgartner}}, \bibinfo {author} {\bibfnamefont
  {P.}~\bibnamefont {Gambardella}}, \bibinfo {author} {\bibfnamefont
  {P.}~\bibnamefont {Němec}}, \bibinfo {author} {\bibfnamefont
  {J.}~\bibnamefont {Wunderlich}}, \bibinfo {author} {\bibfnamefont
  {J.}~\bibnamefont {Sinova}}, \bibinfo {author} {\bibfnamefont
  {P.}~\bibnamefont {Kužel}}, \bibinfo {author} {\bibfnamefont
  {M.}~\bibnamefont {Müller}}, \bibinfo {author} {\bibfnamefont
  {T.}~\bibnamefont {Kampfrath}}, \ and\ \bibinfo {author} {\bibfnamefont
  {T.}~\bibnamefont {Jungwirth}},\ }\href {\doibase 10.1126/sciadv.aar3566}
  {\bibfield  {journal} {\bibinfo  {journal} {Science Advances}\ }\textbf
  {\bibinfo {volume} {4}},\ \bibinfo {pages} {eaar3566} (\bibinfo {year}
  {2018})}\BibitemShut {NoStop}%
\bibitem [{\citenamefont {Roy}\ \emph {et~al.}(2016)\citenamefont {Roy},
  \citenamefont {Otxoa},\ and\ \citenamefont {Wunderlich}}]{Roy2016}%
  \BibitemOpen
  \bibfield  {author} {\bibinfo {author} {\bibfnamefont {P.~E.}\ \bibnamefont
  {Roy}}, \bibinfo {author} {\bibfnamefont {R.~M.}\ \bibnamefont {Otxoa}}, \
  and\ \bibinfo {author} {\bibfnamefont {J.}~\bibnamefont {Wunderlich}},\
  }\href {\doibase 10.1103/PhysRevB.94.014439} {\bibfield  {journal} {\bibinfo
  {journal} {Phys. Rev. B}\ }\textbf {\bibinfo {volume} {94}},\ \bibinfo
  {pages} {014439} (\bibinfo {year} {2016})}\BibitemShut {NoStop}%
\bibitem [{\citenamefont {Bodnar}\ \emph {et~al.}(2018)\citenamefont {Bodnar},
  \citenamefont {{\v{S}}mejkal}, \citenamefont {Turek}, \citenamefont
  {Jungwirth}, \citenamefont {Gomonay}, \citenamefont {Sinova}, \citenamefont
  {Sapozhnik}, \citenamefont {Elmers}, \citenamefont {Kl{\"a}ui},\ and\
  \citenamefont {Jourdan}}]{Bodnar2018}%
  \BibitemOpen
  \bibfield  {author} {\bibinfo {author} {\bibfnamefont {S.~Y.}\ \bibnamefont
  {Bodnar}}, \bibinfo {author} {\bibfnamefont {L.}~\bibnamefont
  {{\v{S}}mejkal}}, \bibinfo {author} {\bibfnamefont {I.}~\bibnamefont
  {Turek}}, \bibinfo {author} {\bibfnamefont {T.}~\bibnamefont {Jungwirth}},
  \bibinfo {author} {\bibfnamefont {O.}~\bibnamefont {Gomonay}}, \bibinfo
  {author} {\bibfnamefont {J.}~\bibnamefont {Sinova}}, \bibinfo {author}
  {\bibfnamefont {A.~A.}\ \bibnamefont {Sapozhnik}}, \bibinfo {author}
  {\bibfnamefont {H.-J.}\ \bibnamefont {Elmers}}, \bibinfo {author}
  {\bibfnamefont {M.}~\bibnamefont {Kl{\"a}ui}}, \ and\ \bibinfo {author}
  {\bibfnamefont {M.}~\bibnamefont {Jourdan}},\ }\href {\doibase
  10.1038/s41467-017-02780-x} {\bibfield  {journal} {\bibinfo  {journal}
  {Nature Communications}\ }\textbf {\bibinfo {volume} {9}},\ \bibinfo {pages}
  {348} (\bibinfo {year} {2018})}\BibitemShut {NoStop}%
\bibitem [{\citenamefont {Meinert}\ \emph {et~al.}(2018)\citenamefont
  {Meinert}, \citenamefont {Graulich},\ and\ \citenamefont
  {Matalla-Wagner}}]{Meinert2018}%
  \BibitemOpen
  \bibfield  {author} {\bibinfo {author} {\bibfnamefont {M.}~\bibnamefont
  {Meinert}}, \bibinfo {author} {\bibfnamefont {D.}~\bibnamefont {Graulich}}, \
  and\ \bibinfo {author} {\bibfnamefont {T.}~\bibnamefont {Matalla-Wagner}},\
  }\href {\doibase 10.1103/PhysRevApplied.9.064040} {\bibfield  {journal}
  {\bibinfo  {journal} {Phys. Rev. Appl.}\ }\textbf {\bibinfo {volume} {9}},\
  \bibinfo {pages} {064040} (\bibinfo {year} {2018})}\BibitemShut {NoStop}%
\bibitem [{\citenamefont {Salemi}\ \emph {et~al.}(2019)\citenamefont {Salemi},
  \citenamefont {Berritta}, \citenamefont {Nandy},\ and\ \citenamefont
  {Oppeneer}}]{Salemi2019}%
  \BibitemOpen
  \bibfield  {author} {\bibinfo {author} {\bibfnamefont {L.}~\bibnamefont
  {Salemi}}, \bibinfo {author} {\bibfnamefont {M.}~\bibnamefont {Berritta}},
  \bibinfo {author} {\bibfnamefont {A.~K.}\ \bibnamefont {Nandy}}, \ and\
  \bibinfo {author} {\bibfnamefont {P.~M.}\ \bibnamefont {Oppeneer}},\ }\href
  {\doibase 10.1038/s41467-019-13367-z} {\bibfield  {journal} {\bibinfo
  {journal} {Nature Communications}\ }\textbf {\bibinfo {volume} {10}},\
  \bibinfo {pages} {5381} (\bibinfo {year} {2019})}\BibitemShut {NoStop}%
\bibitem [{\citenamefont {Selzer}\ \emph {et~al.}(2022)\citenamefont {Selzer},
  \citenamefont {Salemi}, \citenamefont {De\'ak}, \citenamefont {Simon},
  \citenamefont {Szunyogh}, \citenamefont {Oppeneer},\ and\ \citenamefont
  {Nowak}}]{Selzer2022}%
  \BibitemOpen
  \bibfield  {author} {\bibinfo {author} {\bibfnamefont {S.}~\bibnamefont
  {Selzer}}, \bibinfo {author} {\bibfnamefont {L.}~\bibnamefont {Salemi}},
  \bibinfo {author} {\bibfnamefont {A.}~\bibnamefont {De\'ak}}, \bibinfo
  {author} {\bibfnamefont {E.}~\bibnamefont {Simon}}, \bibinfo {author}
  {\bibfnamefont {L.}~\bibnamefont {Szunyogh}}, \bibinfo {author}
  {\bibfnamefont {P.~M.}\ \bibnamefont {Oppeneer}}, \ and\ \bibinfo {author}
  {\bibfnamefont {U.}~\bibnamefont {Nowak}},\ }\href {\doibase
  10.1103/PhysRevB.105.174416} {\bibfield  {journal} {\bibinfo  {journal}
  {Phys. Rev. B}\ }\textbf {\bibinfo {volume} {105}},\ \bibinfo {pages}
  {174416} (\bibinfo {year} {2022})}\BibitemShut {NoStop}%
\bibitem [{\citenamefont {Battiato}\ \emph {et~al.}(2010)\citenamefont
  {Battiato}, \citenamefont {Carva},\ and\ \citenamefont
  {Oppeneer}}]{Battiato2010}%
  \BibitemOpen
  \bibfield  {author} {\bibinfo {author} {\bibfnamefont {M.}~\bibnamefont
  {Battiato}}, \bibinfo {author} {\bibfnamefont {K.}~\bibnamefont {Carva}}, \
  and\ \bibinfo {author} {\bibfnamefont {P.~M.}\ \bibnamefont {Oppeneer}},\
  }\href {\doibase 10.1103/PhysRevLett.105.027203} {\bibfield  {journal}
  {\bibinfo  {journal} {Phys. Rev. Lett.}\ }\textbf {\bibinfo {volume} {105}},\
  \bibinfo {pages} {027203} (\bibinfo {year} {2010})}\BibitemShut {NoStop}%
\bibitem [{\citenamefont {Malinowski}\ \emph {et~al.}(2008)\citenamefont
  {Malinowski}, \citenamefont {Dalla~Longa}, \citenamefont {Rietjens},
  \citenamefont {Paluskar}, \citenamefont {Huijink}, \citenamefont {Swagten},\
  and\ \citenamefont {Koopmans}}]{Malinowski2008}%
  \BibitemOpen
  \bibfield  {author} {\bibinfo {author} {\bibfnamefont {G.}~\bibnamefont
  {Malinowski}}, \bibinfo {author} {\bibfnamefont {F.}~\bibnamefont
  {Dalla~Longa}}, \bibinfo {author} {\bibfnamefont {J.~H.~H.}\ \bibnamefont
  {Rietjens}}, \bibinfo {author} {\bibfnamefont {P.~V.}\ \bibnamefont
  {Paluskar}}, \bibinfo {author} {\bibfnamefont {R.}~\bibnamefont {Huijink}},
  \bibinfo {author} {\bibfnamefont {H.~J.~M.}\ \bibnamefont {Swagten}}, \ and\
  \bibinfo {author} {\bibfnamefont {B.}~\bibnamefont {Koopmans}},\ }\href
  {\doibase 10.1038/nphys1092} {\bibfield  {journal} {\bibinfo  {journal}
  {Nature Physics}\ }\textbf {\bibinfo {volume} {4}},\ \bibinfo {pages} {855}
  (\bibinfo {year} {2008})}\BibitemShut {NoStop}%
\bibitem [{\citenamefont {Melnikov}\ \emph {et~al.}(2011)\citenamefont
  {Melnikov}, \citenamefont {Razdolski}, \citenamefont {Wehling}, \citenamefont
  {Papaioannou}, \citenamefont {Roddatis}, \citenamefont {Fumagalli},
  \citenamefont {Aktsipetrov}, \citenamefont {Lichtenstein},\ and\
  \citenamefont {Bovensiepen}}]{Melnikov2011}%
  \BibitemOpen
  \bibfield  {author} {\bibinfo {author} {\bibfnamefont {A.}~\bibnamefont
  {Melnikov}}, \bibinfo {author} {\bibfnamefont {I.}~\bibnamefont {Razdolski}},
  \bibinfo {author} {\bibfnamefont {T.~O.}\ \bibnamefont {Wehling}}, \bibinfo
  {author} {\bibfnamefont {E.~T.}\ \bibnamefont {Papaioannou}}, \bibinfo
  {author} {\bibfnamefont {V.}~\bibnamefont {Roddatis}}, \bibinfo {author}
  {\bibfnamefont {P.}~\bibnamefont {Fumagalli}}, \bibinfo {author}
  {\bibfnamefont {O.}~\bibnamefont {Aktsipetrov}}, \bibinfo {author}
  {\bibfnamefont {A.~I.}\ \bibnamefont {Lichtenstein}}, \ and\ \bibinfo
  {author} {\bibfnamefont {U.}~\bibnamefont {Bovensiepen}},\ }\href {\doibase
  10.1103/PhysRevLett.107.076601} {\bibfield  {journal} {\bibinfo  {journal}
  {Phys. Rev. Lett.}\ }\textbf {\bibinfo {volume} {107}},\ \bibinfo {pages}
  {076601} (\bibinfo {year} {2011})}\BibitemShut {NoStop}%
\bibitem [{\citenamefont {Rudolf}\ \emph {et~al.}(2012)\citenamefont {Rudolf},
  \citenamefont {La-O-Vorakiat}, \citenamefont {Battiato}, \citenamefont
  {Adam}, \citenamefont {Shaw}, \citenamefont {Turgut}, \citenamefont
  {Maldonado}, \citenamefont {Mathias}, \citenamefont {Grychtol}, \citenamefont
  {Nembach}, \citenamefont {Silva}, \citenamefont {Aeschlimann}, \citenamefont
  {Kapteyn}, \citenamefont {Murnane}, \citenamefont {Schneider},\ and\
  \citenamefont {Oppeneer}}]{Rudolf2012}%
  \BibitemOpen
  \bibfield  {author} {\bibinfo {author} {\bibfnamefont {D.}~\bibnamefont
  {Rudolf}}, \bibinfo {author} {\bibfnamefont {C.}~\bibnamefont
  {La-O-Vorakiat}}, \bibinfo {author} {\bibfnamefont {M.}~\bibnamefont
  {Battiato}}, \bibinfo {author} {\bibfnamefont {R.}~\bibnamefont {Adam}},
  \bibinfo {author} {\bibfnamefont {J.~M.}\ \bibnamefont {Shaw}}, \bibinfo
  {author} {\bibfnamefont {E.}~\bibnamefont {Turgut}}, \bibinfo {author}
  {\bibfnamefont {P.}~\bibnamefont {Maldonado}}, \bibinfo {author}
  {\bibfnamefont {S.}~\bibnamefont {Mathias}}, \bibinfo {author} {\bibfnamefont
  {P.}~\bibnamefont {Grychtol}}, \bibinfo {author} {\bibfnamefont {H.~T.}\
  \bibnamefont {Nembach}}, \bibinfo {author} {\bibfnamefont {T.~J.}\
  \bibnamefont {Silva}}, \bibinfo {author} {\bibfnamefont {M.}~\bibnamefont
  {Aeschlimann}}, \bibinfo {author} {\bibfnamefont {H.~C.}\ \bibnamefont
  {Kapteyn}}, \bibinfo {author} {\bibfnamefont {M.~M.}\ \bibnamefont
  {Murnane}}, \bibinfo {author} {\bibfnamefont {C.~M.}\ \bibnamefont
  {Schneider}}, \ and\ \bibinfo {author} {\bibfnamefont {P.~M.}\ \bibnamefont
  {Oppeneer}},\ }\href {\doibase 10.1038/ncomms2029} {\bibfield  {journal}
  {\bibinfo  {journal} {Nature Communications}\ }\textbf {\bibinfo {volume}
  {3}},\ \bibinfo {pages} {1037} (\bibinfo {year} {2012})}\BibitemShut
  {NoStop}%
\bibitem [{\citenamefont {Eschenlohr}\ \emph {et~al.}(2013)\citenamefont
  {Eschenlohr}, \citenamefont {Battiato}, \citenamefont {Maldonado},
  \citenamefont {Pontius}, \citenamefont {Kachel}, \citenamefont {Holldack},
  \citenamefont {Mitzner}, \citenamefont {F{\"o}hlisch}, \citenamefont
  {Oppeneer},\ and\ \citenamefont {Stamm}}]{Eschenlohr2013}%
  \BibitemOpen
  \bibfield  {author} {\bibinfo {author} {\bibfnamefont {A.}~\bibnamefont
  {Eschenlohr}}, \bibinfo {author} {\bibfnamefont {M.}~\bibnamefont
  {Battiato}}, \bibinfo {author} {\bibfnamefont {P.}~\bibnamefont {Maldonado}},
  \bibinfo {author} {\bibfnamefont {N.}~\bibnamefont {Pontius}}, \bibinfo
  {author} {\bibfnamefont {T.}~\bibnamefont {Kachel}}, \bibinfo {author}
  {\bibfnamefont {K.}~\bibnamefont {Holldack}}, \bibinfo {author}
  {\bibfnamefont {R.}~\bibnamefont {Mitzner}}, \bibinfo {author} {\bibfnamefont
  {A.}~\bibnamefont {F{\"o}hlisch}}, \bibinfo {author} {\bibfnamefont {P.~M.}\
  \bibnamefont {Oppeneer}}, \ and\ \bibinfo {author} {\bibfnamefont
  {C.}~\bibnamefont {Stamm}},\ }\href {\doibase 10.1038/nmat3546} {\bibfield
  {journal} {\bibinfo  {journal} {Nature Materials}\ }\textbf {\bibinfo
  {volume} {12}},\ \bibinfo {pages} {332} (\bibinfo {year} {2013})}\BibitemShut
  {NoStop}%
\bibitem [{\citenamefont {Vodungbo}\ \emph {et~al.}(2016)\citenamefont
  {Vodungbo}, \citenamefont {Tudu}, \citenamefont {Perron}, \citenamefont
  {Delaunay}, \citenamefont {M{\"u}ller}, \citenamefont {Berntsen},
  \citenamefont {Gr{\"u}bel}, \citenamefont {Malinowski}, \citenamefont
  {Weier}, \citenamefont {Gautier}, \citenamefont {Lambert}, \citenamefont
  {Zeitoun}, \citenamefont {Gutt}, \citenamefont {Jal}, \citenamefont {Reid},
  \citenamefont {Granitzka}, \citenamefont {Jaouen}, \citenamefont {Dakovski},
  \citenamefont {Moeller}, \citenamefont {Minitti}, \citenamefont {Mitra},
  \citenamefont {Carron}, \citenamefont {Pfau}, \citenamefont {von
  Korff~Schmising}, \citenamefont {Schneider}, \citenamefont {Eisebitt},\ and\
  \citenamefont {L{\"u}ning}}]{Vodungbo2016}%
  \BibitemOpen
  \bibfield  {author} {\bibinfo {author} {\bibfnamefont {B.}~\bibnamefont
  {Vodungbo}}, \bibinfo {author} {\bibfnamefont {B.}~\bibnamefont {Tudu}},
  \bibinfo {author} {\bibfnamefont {J.}~\bibnamefont {Perron}}, \bibinfo
  {author} {\bibfnamefont {R.}~\bibnamefont {Delaunay}}, \bibinfo {author}
  {\bibfnamefont {L.}~\bibnamefont {M{\"u}ller}}, \bibinfo {author}
  {\bibfnamefont {M.~H.}\ \bibnamefont {Berntsen}}, \bibinfo {author}
  {\bibfnamefont {G.}~\bibnamefont {Gr{\"u}bel}}, \bibinfo {author}
  {\bibfnamefont {G.}~\bibnamefont {Malinowski}}, \bibinfo {author}
  {\bibfnamefont {C.}~\bibnamefont {Weier}}, \bibinfo {author} {\bibfnamefont
  {J.}~\bibnamefont {Gautier}}, \bibinfo {author} {\bibfnamefont
  {G.}~\bibnamefont {Lambert}}, \bibinfo {author} {\bibfnamefont
  {P.}~\bibnamefont {Zeitoun}}, \bibinfo {author} {\bibfnamefont
  {C.}~\bibnamefont {Gutt}}, \bibinfo {author} {\bibfnamefont {E.}~\bibnamefont
  {Jal}}, \bibinfo {author} {\bibfnamefont {A.~H.}\ \bibnamefont {Reid}},
  \bibinfo {author} {\bibfnamefont {P.~W.}\ \bibnamefont {Granitzka}}, \bibinfo
  {author} {\bibfnamefont {N.}~\bibnamefont {Jaouen}}, \bibinfo {author}
  {\bibfnamefont {G.~L.}\ \bibnamefont {Dakovski}}, \bibinfo {author}
  {\bibfnamefont {S.}~\bibnamefont {Moeller}}, \bibinfo {author} {\bibfnamefont
  {M.~P.}\ \bibnamefont {Minitti}}, \bibinfo {author} {\bibfnamefont
  {A.}~\bibnamefont {Mitra}}, \bibinfo {author} {\bibfnamefont
  {S.}~\bibnamefont {Carron}}, \bibinfo {author} {\bibfnamefont
  {B.}~\bibnamefont {Pfau}}, \bibinfo {author} {\bibfnamefont {C.}~\bibnamefont
  {von Korff~Schmising}}, \bibinfo {author} {\bibfnamefont {M.}~\bibnamefont
  {Schneider}}, \bibinfo {author} {\bibfnamefont {S.}~\bibnamefont {Eisebitt}},
  \ and\ \bibinfo {author} {\bibfnamefont {J.}~\bibnamefont {L{\"u}ning}},\
  }\href {\doibase 10.1038/srep18970} {\bibfield  {journal} {\bibinfo
  {journal} {Scientific Reports}\ }\textbf {\bibinfo {volume} {6}},\ \bibinfo
  {pages} {18970} (\bibinfo {year} {2016})}\BibitemShut {NoStop}%
\bibitem [{\citenamefont {Bergeard}\ \emph {et~al.}(2016)\citenamefont
  {Bergeard}, \citenamefont {Hehn}, \citenamefont {Mangin}, \citenamefont
  {Lengaigne}, \citenamefont {Montaigne}, \citenamefont {Lalieu}, \citenamefont
  {Koopmans},\ and\ \citenamefont {Malinowski}}]{Bergeard2016}%
  \BibitemOpen
  \bibfield  {author} {\bibinfo {author} {\bibfnamefont {N.}~\bibnamefont
  {Bergeard}}, \bibinfo {author} {\bibfnamefont {M.}~\bibnamefont {Hehn}},
  \bibinfo {author} {\bibfnamefont {S.}~\bibnamefont {Mangin}}, \bibinfo
  {author} {\bibfnamefont {G.}~\bibnamefont {Lengaigne}}, \bibinfo {author}
  {\bibfnamefont {F.}~\bibnamefont {Montaigne}}, \bibinfo {author}
  {\bibfnamefont {M.~L.~M.}\ \bibnamefont {Lalieu}}, \bibinfo {author}
  {\bibfnamefont {B.}~\bibnamefont {Koopmans}}, \ and\ \bibinfo {author}
  {\bibfnamefont {G.}~\bibnamefont {Malinowski}},\ }\href {\doibase
  10.1103/PhysRevLett.117.147203} {\bibfield  {journal} {\bibinfo  {journal}
  {Phys. Rev. Lett.}\ }\textbf {\bibinfo {volume} {117}},\ \bibinfo {pages}
  {147203} (\bibinfo {year} {2016})}\BibitemShut {NoStop}%
\bibitem [{\citenamefont {Xu}\ \emph {et~al.}(2017)\citenamefont {Xu},
  \citenamefont {Deb}, \citenamefont {Malinowski}, \citenamefont {Hehn},
  \citenamefont {Zhao},\ and\ \citenamefont {Mangin}}]{Xu2017}%
  \BibitemOpen
  \bibfield  {author} {\bibinfo {author} {\bibfnamefont {Y.}~\bibnamefont
  {Xu}}, \bibinfo {author} {\bibfnamefont {M.}~\bibnamefont {Deb}}, \bibinfo
  {author} {\bibfnamefont {G.}~\bibnamefont {Malinowski}}, \bibinfo {author}
  {\bibfnamefont {M.}~\bibnamefont {Hehn}}, \bibinfo {author} {\bibfnamefont
  {W.}~\bibnamefont {Zhao}}, \ and\ \bibinfo {author} {\bibfnamefont
  {S.}~\bibnamefont {Mangin}},\ }\href {\doibase
  https://doi.org/10.1002/adma.201703474} {\bibfield  {journal} {\bibinfo
  {journal} {Advanced Materials}\ }\textbf {\bibinfo {volume} {29}},\ \bibinfo
  {pages} {1703474} (\bibinfo {year} {2017})}\BibitemShut {NoStop}%
\bibitem [{\citenamefont {Alekhin}\ \emph {et~al.}(2017)\citenamefont
  {Alekhin}, \citenamefont {Razdolski}, \citenamefont {Ilin}, \citenamefont
  {Meyburg}, \citenamefont {Diesing}, \citenamefont {Roddatis}, \citenamefont
  {Rungger}, \citenamefont {Stamenova}, \citenamefont {Sanvito}, \citenamefont
  {Bovensiepen},\ and\ \citenamefont {Melnikov}}]{Alekhin2017}%
  \BibitemOpen
  \bibfield  {author} {\bibinfo {author} {\bibfnamefont {A.}~\bibnamefont
  {Alekhin}}, \bibinfo {author} {\bibfnamefont {I.}~\bibnamefont {Razdolski}},
  \bibinfo {author} {\bibfnamefont {N.}~\bibnamefont {Ilin}}, \bibinfo {author}
  {\bibfnamefont {J.~P.}\ \bibnamefont {Meyburg}}, \bibinfo {author}
  {\bibfnamefont {D.}~\bibnamefont {Diesing}}, \bibinfo {author} {\bibfnamefont
  {V.}~\bibnamefont {Roddatis}}, \bibinfo {author} {\bibfnamefont
  {I.}~\bibnamefont {Rungger}}, \bibinfo {author} {\bibfnamefont
  {M.}~\bibnamefont {Stamenova}}, \bibinfo {author} {\bibfnamefont
  {S.}~\bibnamefont {Sanvito}}, \bibinfo {author} {\bibfnamefont
  {U.}~\bibnamefont {Bovensiepen}}, \ and\ \bibinfo {author} {\bibfnamefont
  {A.}~\bibnamefont {Melnikov}},\ }\href {\doibase
  10.1103/PhysRevLett.119.017202} {\bibfield  {journal} {\bibinfo  {journal}
  {Phys. Rev. Lett.}\ }\textbf {\bibinfo {volume} {119}},\ \bibinfo {pages}
  {017202} (\bibinfo {year} {2017})}\BibitemShut {NoStop}%
\bibitem [{\citenamefont {Schellekens}\ \emph {et~al.}(2014)\citenamefont
  {Schellekens}, \citenamefont {Kuiper}, \citenamefont {de~Wit},\ and\
  \citenamefont {Koopmans}}]{Schellekens2014}%
  \BibitemOpen
  \bibfield  {author} {\bibinfo {author} {\bibfnamefont {A.~J.}\ \bibnamefont
  {Schellekens}}, \bibinfo {author} {\bibfnamefont {K.~C.}\ \bibnamefont
  {Kuiper}}, \bibinfo {author} {\bibfnamefont {R.~R. J.~C.}\ \bibnamefont
  {de~Wit}}, \ and\ \bibinfo {author} {\bibfnamefont {B.}~\bibnamefont
  {Koopmans}},\ }\href {\doibase 10.1038/ncomms5333} {\bibfield  {journal}
  {\bibinfo  {journal} {Nature Communications}\ }\textbf {\bibinfo {volume}
  {5}},\ \bibinfo {pages} {4333} (\bibinfo {year} {2014})}\BibitemShut
  {NoStop}%
\bibitem [{\citenamefont {Choi}\ \emph {et~al.}(2014)\citenamefont {Choi},
  \citenamefont {Min}, \citenamefont {Lee},\ and\ \citenamefont
  {Cahill}}]{Choi2014}%
  \BibitemOpen
  \bibfield  {author} {\bibinfo {author} {\bibfnamefont {G.-M.}\ \bibnamefont
  {Choi}}, \bibinfo {author} {\bibfnamefont {B.-C.}\ \bibnamefont {Min}},
  \bibinfo {author} {\bibfnamefont {K.-J.}\ \bibnamefont {Lee}}, \ and\
  \bibinfo {author} {\bibfnamefont {D.~G.}\ \bibnamefont {Cahill}},\ }\href
  {\doibase 10.1038/ncomms5334} {\bibfield  {journal} {\bibinfo  {journal}
  {Nature Communications}\ }\textbf {\bibinfo {volume} {5}},\ \bibinfo {pages}
  {4334} (\bibinfo {year} {2014})}\BibitemShut {NoStop}%
\bibitem [{\citenamefont {Razdolski}\ \emph {et~al.}(2017)\citenamefont
  {Razdolski}, \citenamefont {Alekhin}, \citenamefont {Ilin}, \citenamefont
  {Meyburg}, \citenamefont {Roddatis}, \citenamefont {Diesing}, \citenamefont
  {Bovensiepen},\ and\ \citenamefont {Melnikov}}]{Razdolski2017}%
  \BibitemOpen
  \bibfield  {author} {\bibinfo {author} {\bibfnamefont {I.}~\bibnamefont
  {Razdolski}}, \bibinfo {author} {\bibfnamefont {A.}~\bibnamefont {Alekhin}},
  \bibinfo {author} {\bibfnamefont {N.}~\bibnamefont {Ilin}}, \bibinfo {author}
  {\bibfnamefont {J.~P.}\ \bibnamefont {Meyburg}}, \bibinfo {author}
  {\bibfnamefont {V.}~\bibnamefont {Roddatis}}, \bibinfo {author}
  {\bibfnamefont {D.}~\bibnamefont {Diesing}}, \bibinfo {author} {\bibfnamefont
  {U.}~\bibnamefont {Bovensiepen}}, \ and\ \bibinfo {author} {\bibfnamefont
  {A.}~\bibnamefont {Melnikov}},\ }\href {\doibase 10.1038/ncomms15007}
  {\bibfield  {journal} {\bibinfo  {journal} {Nature Communications}\ }\textbf
  {\bibinfo {volume} {8}},\ \bibinfo {pages} {15007} (\bibinfo {year}
  {2017})}\BibitemShut {NoStop}%
\bibitem [{\citenamefont {Ritzmann}\ \emph {et~al.}(2020)\citenamefont
  {Ritzmann}, \citenamefont {Bal\'a\ifmmode~\check{z}\else \v{z}\fi{}},
  \citenamefont {Maldonado}, \citenamefont {Carva},\ and\ \citenamefont
  {Oppeneer}}]{Ritzmann2020}%
  \BibitemOpen
  \bibfield  {author} {\bibinfo {author} {\bibfnamefont {U.}~\bibnamefont
  {Ritzmann}}, \bibinfo {author} {\bibfnamefont {P.}~\bibnamefont
  {Bal\'a\ifmmode~\check{z}\else \v{z}\fi{}}}, \bibinfo {author} {\bibfnamefont
  {P.}~\bibnamefont {Maldonado}}, \bibinfo {author} {\bibfnamefont
  {K.}~\bibnamefont {Carva}}, \ and\ \bibinfo {author} {\bibfnamefont {P.~M.}\
  \bibnamefont {Oppeneer}},\ }\href {\doibase 10.1103/PhysRevB.101.174427}
  {\bibfield  {journal} {\bibinfo  {journal} {Phys. Rev. B}\ }\textbf {\bibinfo
  {volume} {101}},\ \bibinfo {pages} {174427} (\bibinfo {year}
  {2020})}\BibitemShut {NoStop}%
\bibitem [{\citenamefont {Gomonay}\ and\ \citenamefont
  {Loktev}(2010)}]{Gomonay2010}%
  \BibitemOpen
  \bibfield  {author} {\bibinfo {author} {\bibfnamefont {H.~V.}\ \bibnamefont
  {Gomonay}}\ and\ \bibinfo {author} {\bibfnamefont {V.~M.}\ \bibnamefont
  {Loktev}},\ }\href {\doibase 10.1103/PhysRevB.81.144427} {\bibfield
  {journal} {\bibinfo  {journal} {Phys. Rev. B}\ }\textbf {\bibinfo {volume}
  {81}},\ \bibinfo {pages} {144427} (\bibinfo {year} {2010})}\BibitemShut
  {NoStop}%
\bibitem [{\citenamefont {Chirac}\ \emph {et~al.}(2020)\citenamefont {Chirac},
  \citenamefont {Chauleau}, \citenamefont {Thibaudeau}, \citenamefont
  {Gomonay},\ and\ \citenamefont {Viret}}]{Chirac2020}%
  \BibitemOpen
  \bibfield  {author} {\bibinfo {author} {\bibfnamefont {T.}~\bibnamefont
  {Chirac}}, \bibinfo {author} {\bibfnamefont {J.-Y.}\ \bibnamefont
  {Chauleau}}, \bibinfo {author} {\bibfnamefont {P.}~\bibnamefont
  {Thibaudeau}}, \bibinfo {author} {\bibfnamefont {O.}~\bibnamefont {Gomonay}},
  \ and\ \bibinfo {author} {\bibfnamefont {M.}~\bibnamefont {Viret}},\ }\href
  {\doibase 10.1103/PhysRevB.102.134415} {\bibfield  {journal} {\bibinfo
  {journal} {Phys. Rev. B}\ }\textbf {\bibinfo {volume} {102}},\ \bibinfo
  {pages} {134415} (\bibinfo {year} {2020})}\BibitemShut {NoStop}%
\bibitem [{\citenamefont {Nowak}(2007)}]{Nowak2007}%
  \BibitemOpen
  \bibfield  {author} {\bibinfo {author} {\bibfnamefont {U.}~\bibnamefont
  {Nowak}},\ }\enquote {\bibinfo {title} {Classical spin models},}\ in\ \href
  {\doibase 10.1002/97804470022184.hmm205} {\emph {\bibinfo {booktitle}
  {Handbook of Magnetism and Advanced Magnetic Materials}}},\ \bibinfo {editor}
  {edited by\ \bibinfo {editor} {\bibfnamefont {H.}~\bibnamefont
  {Kronm{\"u}ller}}\ and\ \bibinfo {editor} {\bibfnamefont {S.}~\bibnamefont
  {Parkin}}}\ (\bibinfo  {publisher} {J. Wiley \& Sons, New York},\ \bibinfo
  {year} {2007})\ pp.\ \bibinfo {pages} {858--876}\BibitemShut {NoStop}%
\bibitem [{\citenamefont {Battiato}\ \emph {et~al.}(2012)\citenamefont
  {Battiato}, \citenamefont {Carva},\ and\ \citenamefont
  {Oppeneer}}]{Battiato2012}%
  \BibitemOpen
  \bibfield  {author} {\bibinfo {author} {\bibfnamefont {M.}~\bibnamefont
  {Battiato}}, \bibinfo {author} {\bibfnamefont {K.}~\bibnamefont {Carva}}, \
  and\ \bibinfo {author} {\bibfnamefont {P.~M.}\ \bibnamefont {Oppeneer}},\
  }\href {\doibase 10.1103/PhysRevB.86.024404} {\bibfield  {journal} {\bibinfo
  {journal} {Phys. Rev. B}\ }\textbf {\bibinfo {volume} {86}},\ \bibinfo
  {pages} {024404} (\bibinfo {year} {2012})}\BibitemShut {NoStop}%
\bibitem [{Sup()}]{SupplMat}%
  \BibitemOpen
  \href@noop {} {}\bibinfo {note} {See Supplemental Material for discussions of
  (i) the superdiffusive spin-transport model, (ii) the spin wave dispersion in
  Mn$_2$Au and its relation to the frequency spectra, (iii) the impact of the
  FWHM and the penetration depth on switching and (iv) the impact of
  temperature on switching.}\BibitemShut {Stop}%
\bibitem [{\citenamefont {Lu}\ \emph {et~al.}(2020)\citenamefont {Lu},
  \citenamefont {Zhao}, \citenamefont {Battiato}, \citenamefont {Wu},\ and\
  \citenamefont {Yuan}}]{Lu2020}%
  \BibitemOpen
  \bibfield  {author} {\bibinfo {author} {\bibfnamefont {W.-T.}\ \bibnamefont
  {Lu}}, \bibinfo {author} {\bibfnamefont {Y.}~\bibnamefont {Zhao}}, \bibinfo
  {author} {\bibfnamefont {M.}~\bibnamefont {Battiato}}, \bibinfo {author}
  {\bibfnamefont {Y.}~\bibnamefont {Wu}}, \ and\ \bibinfo {author}
  {\bibfnamefont {Z.}~\bibnamefont {Yuan}},\ }\href {\doibase
  10.1103/PhysRevB.101.014435} {\bibfield  {journal} {\bibinfo  {journal}
  {Phys. Rev. B}\ }\textbf {\bibinfo {volume} {101}},\ \bibinfo {pages}
  {014435} (\bibinfo {year} {2020})}\BibitemShut {NoStop}%
\bibitem [{Note1()}]{Note1}%
  \BibitemOpen
  \bibinfo {note} {The FWHM of a Gaussian pulse is related to its standard
  deviation via $\protect \mathrm {FWHM}\approx 2.355 \Delta $}\BibitemShut
  {NoStop}%
\bibitem [{\citenamefont {Slonczewski}(1996)}]{Slonzweski1996}%
  \BibitemOpen
  \bibfield  {author} {\bibinfo {author} {\bibfnamefont {J.}~\bibnamefont
  {Slonczewski}},\ }\href {\doibase
  https://doi.org/10.1016/0304-8853(96)00062-5} {\bibfield  {journal} {\bibinfo
   {journal} {Journal of Magnetism and Magnetic Materials}\ }\textbf {\bibinfo
  {volume} {159}},\ \bibinfo {pages} {L1} (\bibinfo {year} {1996})}\BibitemShut
  {NoStop}%
\bibitem [{\citenamefont {Berger}(1996)}]{Berger1996}%
  \BibitemOpen
  \bibfield  {author} {\bibinfo {author} {\bibfnamefont {L.}~\bibnamefont
  {Berger}},\ }\href {\doibase 10.1103/PhysRevB.54.9353} {\bibfield  {journal}
  {\bibinfo  {journal} {Phys. Rev. B}\ }\textbf {\bibinfo {volume} {54}},\
  \bibinfo {pages} {9353} (\bibinfo {year} {1996})}\BibitemShut {NoStop}%
\bibitem [{\citenamefont {Slonczewski}(2002)}]{Slonzweski2002}%
  \BibitemOpen
  \bibfield  {author} {\bibinfo {author} {\bibfnamefont {J.~C.}\ \bibnamefont
  {Slonczewski}},\ }\href {\doibase
  https://doi.org/10.1016/S0304-8853(02)00291-3} {\bibfield  {journal}
  {\bibinfo  {journal} {Journal of Magnetism and Magnetic Materials}\ }\textbf
  {\bibinfo {volume} {247}},\ \bibinfo {pages} {324} (\bibinfo {year}
  {2002})}\BibitemShut {NoStop}%
\bibitem [{\citenamefont {Kazantseva}\ \emph {et~al.}(2008)\citenamefont
  {Kazantseva}, \citenamefont {Hinzke}, \citenamefont {Nowak}, \citenamefont
  {Chantrell}, \citenamefont {Atxitia},\ and\ \citenamefont
  {Chubykalo-Fesenko}}]{Kazantseva2008}%
  \BibitemOpen
  \bibfield  {author} {\bibinfo {author} {\bibfnamefont {N.}~\bibnamefont
  {Kazantseva}}, \bibinfo {author} {\bibfnamefont {D.}~\bibnamefont {Hinzke}},
  \bibinfo {author} {\bibfnamefont {U.}~\bibnamefont {Nowak}}, \bibinfo
  {author} {\bibfnamefont {R.~W.}\ \bibnamefont {Chantrell}}, \bibinfo {author}
  {\bibfnamefont {U.}~\bibnamefont {Atxitia}}, \ and\ \bibinfo {author}
  {\bibfnamefont {O.}~\bibnamefont {Chubykalo-Fesenko}},\ }\href {\doibase
  10.1103/PhysRevB.77.184428} {\bibfield  {journal} {\bibinfo  {journal} {Phys.
  Rev. B}\ }\textbf {\bibinfo {volume} {77}},\ \bibinfo {pages} {184428}
  (\bibinfo {year} {2008})}\BibitemShut {NoStop}%
\bibitem [{\citenamefont {Baláž}\ \emph {et~al.}(2018)\citenamefont
  {Baláž}, \citenamefont {Žonda}, \citenamefont {Carva}, \citenamefont
  {Maldonado},\ and\ \citenamefont {Oppeneer}}]{Balaz2018}%
  \BibitemOpen
  \bibfield  {author} {\bibinfo {author} {\bibfnamefont {P.}~\bibnamefont
  {Baláž}}, \bibinfo {author} {\bibfnamefont {M.}~\bibnamefont {Žonda}},
  \bibinfo {author} {\bibfnamefont {K.}~\bibnamefont {Carva}}, \bibinfo
  {author} {\bibfnamefont {P.}~\bibnamefont {Maldonado}}, \ and\ \bibinfo
  {author} {\bibfnamefont {P.~M.}\ \bibnamefont {Oppeneer}},\ }\href {\doibase
  10.1088/1361-648X/aaad95} {\bibfield  {journal} {\bibinfo  {journal} {Journal
  of Physics: Condensed Matter}\ }\textbf {\bibinfo {volume} {30}},\ \bibinfo
  {pages} {115801} (\bibinfo {year} {2018})}\BibitemShut {NoStop}%
\bibitem [{\citenamefont {Stiles}\ and\ \citenamefont
  {Zangwill}(2002)}]{Stiles2002}%
  \BibitemOpen
  \bibfield  {author} {\bibinfo {author} {\bibfnamefont {M.~D.}\ \bibnamefont
  {Stiles}}\ and\ \bibinfo {author} {\bibfnamefont {A.}~\bibnamefont
  {Zangwill}},\ }\href {\doibase 10.1103/PhysRevB.66.014407} {\bibfield
  {journal} {\bibinfo  {journal} {Phys. Rev. B}\ }\textbf {\bibinfo {volume}
  {66}},\ \bibinfo {pages} {014407} (\bibinfo {year} {2002})}\BibitemShut
  {NoStop}%
\bibitem [{\citenamefont {Ghosh}\ \emph {et~al.}(2012)\citenamefont {Ghosh},
  \citenamefont {Auffret}, \citenamefont {Ebels},\ and\ \citenamefont
  {Bailey}}]{Ghosh2012}%
  \BibitemOpen
  \bibfield  {author} {\bibinfo {author} {\bibfnamefont {A.}~\bibnamefont
  {Ghosh}}, \bibinfo {author} {\bibfnamefont {S.}~\bibnamefont {Auffret}},
  \bibinfo {author} {\bibfnamefont {U.}~\bibnamefont {Ebels}}, \ and\ \bibinfo
  {author} {\bibfnamefont {W.~E.}\ \bibnamefont {Bailey}},\ }\href {\doibase
  10.1103/PhysRevLett.109.127202} {\bibfield  {journal} {\bibinfo  {journal}
  {Phys. Rev. Lett.}\ }\textbf {\bibinfo {volume} {109}},\ \bibinfo {pages}
  {127202} (\bibinfo {year} {2012})}\BibitemShut {NoStop}%
\bibitem [{not()}]{note}%
  \BibitemOpen
  \href@noop {} {}\bibinfo {note} {The numerical integration of the \ac{LLG}
  equation is performed using the Heun method \cite{Nowak2007} with a timestep
  $\Delta t$ of $\SI{0.33}{\femto\second}$. We consider a cross section of
  $2\times \SI{2}{\nano\metre\squared}$ with periodic boundary conditions along
  $x$ and $y$ directions of the Mn$_2$Au layer.}\BibitemShut {Stop}%
\bibitem [{\citenamefont {Barthem}\ \emph {et~al.}(2013)\citenamefont
  {Barthem}, \citenamefont {Colin}, \citenamefont {Mayaffre}, \citenamefont
  {Julien},\ and\ \citenamefont {Givord}}]{Barthem2013}%
  \BibitemOpen
  \bibfield  {author} {\bibinfo {author} {\bibfnamefont {V.~M. T.~S.}\
  \bibnamefont {Barthem}}, \bibinfo {author} {\bibfnamefont {C.~V.}\
  \bibnamefont {Colin}}, \bibinfo {author} {\bibfnamefont {H.}~\bibnamefont
  {Mayaffre}}, \bibinfo {author} {\bibfnamefont {M.-H.}\ \bibnamefont
  {Julien}}, \ and\ \bibinfo {author} {\bibfnamefont {D.}~\bibnamefont
  {Givord}},\ }\href {\doibase 10.1038/ncomms3892} {\bibfield  {journal}
  {\bibinfo  {journal} {Nature Communications}\ }\textbf {\bibinfo {volume}
  {4}},\ \bibinfo {pages} {2892} (\bibinfo {year} {2013})}\BibitemShut
  {NoStop}%
\bibitem [{\citenamefont {Vaidya}\ \emph {et~al.}(2020)\citenamefont {Vaidya},
  \citenamefont {Morley}, \citenamefont {van Tol}, \citenamefont {Liu},
  \citenamefont {Cheng}, \citenamefont {Brataas}, \citenamefont {Lederman},\
  and\ \citenamefont {del Barco}}]{Vaidya2020}%
  \BibitemOpen
  \bibfield  {author} {\bibinfo {author} {\bibfnamefont {P.}~\bibnamefont
  {Vaidya}}, \bibinfo {author} {\bibfnamefont {S.~A.}\ \bibnamefont {Morley}},
  \bibinfo {author} {\bibfnamefont {J.}~\bibnamefont {van Tol}}, \bibinfo
  {author} {\bibfnamefont {Y.}~\bibnamefont {Liu}}, \bibinfo {author}
  {\bibfnamefont {R.}~\bibnamefont {Cheng}}, \bibinfo {author} {\bibfnamefont
  {A.}~\bibnamefont {Brataas}}, \bibinfo {author} {\bibfnamefont
  {D.}~\bibnamefont {Lederman}}, \ and\ \bibinfo {author} {\bibfnamefont
  {E.}~\bibnamefont {del Barco}},\ }\href {\doibase 10.1126/science.aaz4247}
  {\bibfield  {journal} {\bibinfo  {journal} {Science}\ }\textbf {\bibinfo
  {volume} {368}},\ \bibinfo {pages} {160} (\bibinfo {year}
  {2020})}\BibitemShut {NoStop}%
\bibitem [{\citenamefont {Li}\ \emph {et~al.}(2020)\citenamefont {Li},
  \citenamefont {Wilson}, \citenamefont {Cheng}, \citenamefont {Lohmann},
  \citenamefont {Kavand}, \citenamefont {Yuan}, \citenamefont {Aldosary},
  \citenamefont {Agladze}, \citenamefont {Wei}, \citenamefont {Sherwin},\ and\
  \citenamefont {Shi}}]{Li2020}%
  \BibitemOpen
  \bibfield  {author} {\bibinfo {author} {\bibfnamefont {J.}~\bibnamefont
  {Li}}, \bibinfo {author} {\bibfnamefont {C.~B.}\ \bibnamefont {Wilson}},
  \bibinfo {author} {\bibfnamefont {R.}~\bibnamefont {Cheng}}, \bibinfo
  {author} {\bibfnamefont {M.}~\bibnamefont {Lohmann}}, \bibinfo {author}
  {\bibfnamefont {M.}~\bibnamefont {Kavand}}, \bibinfo {author} {\bibfnamefont
  {W.}~\bibnamefont {Yuan}}, \bibinfo {author} {\bibfnamefont {M.}~\bibnamefont
  {Aldosary}}, \bibinfo {author} {\bibfnamefont {N.}~\bibnamefont {Agladze}},
  \bibinfo {author} {\bibfnamefont {P.}~\bibnamefont {Wei}}, \bibinfo {author}
  {\bibfnamefont {M.~S.}\ \bibnamefont {Sherwin}}, \ and\ \bibinfo {author}
  {\bibfnamefont {J.}~\bibnamefont {Shi}},\ }\href {\doibase
  10.1038/s41586-020-1950-4} {\bibfield  {journal} {\bibinfo  {journal}
  {Nature}\ }\textbf {\bibinfo {volume} {578}},\ \bibinfo {pages} {70}
  (\bibinfo {year} {2020})}\BibitemShut {NoStop}%
\bibitem [{\citenamefont {Salikhov}\ \emph {et~al.}(2023)\citenamefont
  {Salikhov}, \citenamefont {Ilyakov}, \citenamefont {K{\"o}rber},
  \citenamefont {K{\'a}kay}, \citenamefont {Gallardo}, \citenamefont
  {Ponomaryov}, \citenamefont {Deinert}, \citenamefont {de~Oliveira},
  \citenamefont {Lenz}, \citenamefont {Fassbender}, \citenamefont {Bonetti},
  \citenamefont {Hellwig}, \citenamefont {Lindner},\ and\ \citenamefont
  {Kovalev}}]{Salikhov2023}%
  \BibitemOpen
  \bibfield  {author} {\bibinfo {author} {\bibfnamefont {R.}~\bibnamefont
  {Salikhov}}, \bibinfo {author} {\bibfnamefont {I.}~\bibnamefont {Ilyakov}},
  \bibinfo {author} {\bibfnamefont {L.}~\bibnamefont {K{\"o}rber}}, \bibinfo
  {author} {\bibfnamefont {A.}~\bibnamefont {K{\'a}kay}}, \bibinfo {author}
  {\bibfnamefont {R.~A.}\ \bibnamefont {Gallardo}}, \bibinfo {author}
  {\bibfnamefont {A.}~\bibnamefont {Ponomaryov}}, \bibinfo {author}
  {\bibfnamefont {J.-C.}\ \bibnamefont {Deinert}}, \bibinfo {author}
  {\bibfnamefont {T.~V. A.~G.}\ \bibnamefont {de~Oliveira}}, \bibinfo {author}
  {\bibfnamefont {K.}~\bibnamefont {Lenz}}, \bibinfo {author} {\bibfnamefont
  {J.}~\bibnamefont {Fassbender}}, \bibinfo {author} {\bibfnamefont
  {S.}~\bibnamefont {Bonetti}}, \bibinfo {author} {\bibfnamefont
  {O.}~\bibnamefont {Hellwig}}, \bibinfo {author} {\bibfnamefont
  {J.}~\bibnamefont {Lindner}}, \ and\ \bibinfo {author} {\bibfnamefont
  {S.}~\bibnamefont {Kovalev}},\ }\href {\doibase 10.1038/s41567-022-01908-1}
  {\bibfield  {journal} {\bibinfo  {journal} {Nature Physics}\ } (\bibinfo
  {year} {2023}),\ 10.1038/s41567-022-01908-1}\BibitemShut {NoStop}%
\bibitem [{\citenamefont {Rezende}\ \emph {et~al.}(2019)\citenamefont
  {Rezende}, \citenamefont {Azevedo},\ and\ \citenamefont
  {Rodríguez-Suárez}}]{Rezende2019}%
  \BibitemOpen
  \bibfield  {author} {\bibinfo {author} {\bibfnamefont {S.~M.}\ \bibnamefont
  {Rezende}}, \bibinfo {author} {\bibfnamefont {A.}~\bibnamefont {Azevedo}}, \
  and\ \bibinfo {author} {\bibfnamefont {R.~L.}\ \bibnamefont
  {Rodríguez-Suárez}},\ }\href {\doibase 10.1063/1.5109132} {\bibfield
  {journal} {\bibinfo  {journal} {Journal of Applied Physics}\ }\textbf
  {\bibinfo {volume} {126}},\ \bibinfo {pages} {151101} (\bibinfo {year}
  {2019})}\BibitemShut {NoStop}%
\bibitem [{\citenamefont {Kittel}(1951)}]{Kittel1951}%
  \BibitemOpen
  \bibfield  {author} {\bibinfo {author} {\bibfnamefont {C.}~\bibnamefont
  {Kittel}},\ }\href {\doibase 10.1103/PhysRev.82.565} {\bibfield  {journal}
  {\bibinfo  {journal} {Phys. Rev.}\ }\textbf {\bibinfo {volume} {82}},\
  \bibinfo {pages} {565} (\bibinfo {year} {1951})}\BibitemShut {NoStop}%
\bibitem [{\citenamefont {Gomonay}\ and\ \citenamefont
  {Loktev}(2014)}]{Gomonay2014}%
  \BibitemOpen
  \bibfield  {author} {\bibinfo {author} {\bibfnamefont {E.~V.}\ \bibnamefont
  {Gomonay}}\ and\ \bibinfo {author} {\bibfnamefont {V.~M.}\ \bibnamefont
  {Loktev}},\ }\href {\doibase 10.1063/1.4862467} {\bibfield  {journal}
  {\bibinfo  {journal} {Low Temperature Physics}\ }\textbf {\bibinfo {volume}
  {40}},\ \bibinfo {pages} {17} (\bibinfo {year} {2014})}\BibitemShut {NoStop}%
\bibitem [{\citenamefont {Dannegger}\ \emph {et~al.}(2021)\citenamefont
  {Dannegger}, \citenamefont {Berritta}, \citenamefont {Carva}, \citenamefont
  {Selzer}, \citenamefont {Ritzmann}, \citenamefont {Oppeneer},\ and\
  \citenamefont {Nowak}}]{Dannegger2021}%
  \BibitemOpen
  \bibfield  {author} {\bibinfo {author} {\bibfnamefont {T.}~\bibnamefont
  {Dannegger}}, \bibinfo {author} {\bibfnamefont {M.}~\bibnamefont {Berritta}},
  \bibinfo {author} {\bibfnamefont {K.}~\bibnamefont {Carva}}, \bibinfo
  {author} {\bibfnamefont {S.}~\bibnamefont {Selzer}}, \bibinfo {author}
  {\bibfnamefont {U.}~\bibnamefont {Ritzmann}}, \bibinfo {author}
  {\bibfnamefont {P.~M.}\ \bibnamefont {Oppeneer}}, \ and\ \bibinfo {author}
  {\bibfnamefont {U.}~\bibnamefont {Nowak}},\ }\href {\doibase
  10.1103/PhysRevB.104.L060413} {\bibfield  {journal} {\bibinfo  {journal}
  {Phys. Rev. B}\ }\textbf {\bibinfo {volume} {104}},\ \bibinfo {pages}
  {L060413} (\bibinfo {year} {2021})}\BibitemShut {NoStop}%
\end{thebibliography}
%

\end{document}


\title{Supplemental Material: N\'eel-Vector Switching and THz Spin-Wave Excitation 
in Mn$_{2}$Au due to Femtosecond Spin-Transfer Torques}

\author{Markus Weißenhofer}
\email[]{markus.weissenhofer@fu-berlin.de}
 \affiliation{Department of Physics and Astronomy, Uppsala University, P.\ O.\ Box 516, S-751 20 Uppsala, Sweden}
\affiliation{Department of Physics, Freie Universit{\"a}t Berlin, Arnimallee 14, D-14195 Berlin, Germany}

\author{Francesco Foggetti}
\affiliation{Department of Physics and Astronomy, Uppsala University, P.\ O.\ Box 516, S-751 20 Uppsala, Sweden}

\author{Ulrich Nowak}
\affiliation{Department of Physics, University of Konstanz,
D-78457 Konstanz, Germany}

\author{Peter M.\ Oppeneer}
\affiliation{Department of Physics and Astronomy, Uppsala University, P.\ O.\ Box 516, S-751 20 Uppsala, Sweden}

\maketitle

\begin{acronym}
\acro{AFM}[AFM]{antiferromagnetic}
\acro{FM}[FM]{ferromagnetic}
\acro{NM}[NM]{nonmagnetic}
\acro{AFMR}[AFMR]{antiferromagnetic resonance}
\acro{LLG}[LLG]{Landau-Lifshitz-Gilbert}
\acro{LSWT}[LSWT]{linear spin-wave theory}
\acro{STT}[STT]{spin-transfer torque}
\acro{SSW}[SSW]{standing spin wave}
\end{acronym}

\section{Superdiffusive spin-transport model}
The superdiffusive spin-transport model is a semiclassical theory specifically designed to describe the transport of laser-excited non-equilibrium electrons. It was initially developed in an effort to explain ultrafast demagnetization experiments on Ni \cite{Battiato2010} and later extended to describe magnetic heterostructures consisting of a \ac{FM} and a \ac{NM} layer \cite{Battiato2012}.

The model assumes that the laser pulse excites two channels of spin-up and spin-down electrons, respectively, in the \ac{FM} layer. Since in the \ac{FM} layer the two channels have different transport properties (i.e.,\ lifetime and velocities of electrons), the resulting net current is a spin-polarized current which in turn is responsible for the femtosecond demagnetization of the FM layer when the spin-polarized current is injected into the \ac{NM} layer.

The key equation of the superdiffusive spin-transport model is the equation of motion for the excited hot-electron density $n_\sigma(E,z,t)$, where $\sigma$ is the electron spin, $E$ the energy and $z$ the position,
\begin{equation}
    \frac{\partial n_\sigma}{\partial t}+\frac{n_\sigma}{\tau_\sigma}=\left(-\frac{\partial}{\partial t}\hat\phi +\hat I \right)S^{\mathrm{eff}}_\sigma.
    \label{eq:superdifussion}
\end{equation}
For details and the derivation of the above equation, we refer to Ref.~\cite{Battiato2010}. For the sake of simplicity {\red of notation}, the energy, position, and time dependence has been omitted from the equation. In the model, the \ac{FM}/\ac{NM} structure is described as a quasi-1D system, where $z$ is the non-trivial spatial coordinate perpendicular to the layers, $\tau_\sigma$ is an energy dependent quantity that describes the lifetime of electrons (i.e.\ the time between two scattering events), $\hat\phi$ is the flux operator that takes into account the electron dynamics, including lifetimes, velocities and interaction with the interfaces. $\hat I$ is the identity operator and $S^{\mathrm{eff}}_\sigma$ is the effective spin source, a spin, energy, position, and time-dependent quantity that consists of two contributions: the electrons excited from the external laser source $S^{\mathrm{ext}}_\sigma$ and the electrons generated by internal scattering events $S^\mathrm{p}_\sigma$, such that $S^\mathrm{eff}_\sigma=S^\mathrm{ext}_\sigma+S^\mathrm{p}_\sigma$.

 Solving the superdiffusive transport equation gives the electron density $n_\sigma$ as well as the spin-current density $j_\mathrm{s}(z,t)$, which is defined as the difference between spin-up and spin-down electrons flowing in and out at each position $z$ \cite{Lu2020}. 

\section{Dispersion relation of spin waves in Mn$_2$Au}

Spin-wave dispersion relations for arbitrary systems can be efficiently calculated using \ac{LSWT} (see, e.g., Ref.~\cite{KRANENDONK1958}). This approach rests on an expansion of the spin Hamiltonian to second order in the deviations from a reference state, typically the groundstate (or any other metastable state). 

Here, we map the spin Hamiltonian for Mn$_2$Au,
\begin{align}
\mathcal{H}
=
-
\frac{1}{2}
\sum_{i\neq j}
J_{ij}
\vec{S}_i
\cdot
\vec{S}_j
-
d_z
\sum_{i}
{S}_{i,z}^2
-
d_{zz}
\sum_{i}
{S}_{i,z}^4
-
d_{xy}
\sum_{i}
{S}_{i,x}^2
{S}_{i,y}^2,
\label{eq:Hamiltonian}
\end{align}
onto a one-dimensional biaxial antiferromagnet that extends along the $z$-axis and where only an effective nearest-neighbor exchange is considered. We do this because for such a system analytical solutions exist in literature \cite{Rezende2019}. 

In the groundstate, the N{\'e}el vectors of Mn$_2$Au are oriented along the $[110]$ direction (or any other crystallographically equivalent axis). Thus, as a first step, we rotate the coordinate system by $\frac{\pi}{4}$ around the $z$-axis. In this new basis (indicated by ``$\sim$"), the groundstate N{\'e}el vector $\tilde{\vec{S}}_0$ is parallel to a basis vector; without loss of generality we assume that $\tilde{\vec{S}}_0 = (\pm 1,0,0)^\mathrm{T}$. Keeping only terms up to quadratic order in the components orthogonal to the groundstate, which we denote by $\tilde{\vec{S}}_i^\perp$, and assuming homogeneity of the spin configuration within the $x$-$y$-plane, we can map Eq.~\eqref{eq:Hamiltonian} onto an effective one-dimensional spin Hamiltonian that reads
\begin{equation}
\tilde{\mathcal{H}} = 
\tilde{\mathcal{H}}_0
-
J^\mathrm{inter} 
\sum_{i=1}^{N_\mathrm{layer}-1}
\tilde{\vec{S}}_i^\perp  \cdot \tilde{\vec{S}}_{i+1}^\perp
-
d_z
\sum_{i=1}^{N_\mathrm{layer}}
 \tilde{S}_{i,z}^2
-
d_{xy}
\sum_{i=1}^{N_\mathrm{layer}}
 \tilde{S}_{i,y}^2
,
\end{equation}
with $\tilde{\mathcal{H}}_0$ being the irrelevant groundstate energy and $N_\mathrm{layer}$ being the number of Mn layers along the $z$-axis. Note that we have also replaced the long-range exchange interactions with an effective nearest-neighbor exchange. The corresponding exchange constant was calculated in Ref.~\cite{Selzer2022} and has a value $J^\mathrm{inter}=\SI{371.13}{\milli\electronvolt}$.

The two emergent (positive) \ac{AFMR} frequencies for this effective spin Hamiltonian, i.e., the spin-wave eigenmodes for which the wave number $k$ is zero, can be calculated using formulas from the literature \cite{Rezende2019},
\begin{align}
    f_0^\mathrm{a} &= \frac{\gamma}{2 \pi\mu_\mathrm{s}} \sqrt{2J^\mathrm{inter}d_{xy}} \approx \SI{0.85}{\tera\hertz}, \\
    f_0^\mathrm{b} &= \frac{\gamma}{2 \pi\mu_\mathrm{s}} \sqrt{2J^\mathrm{inter}(d_{xy}-d_z)} \approx \SI{2.9}{\tera\hertz}.
\end{align}

\begin{figure}
    \centering
\includegraphics[width=0.5\textwidth]{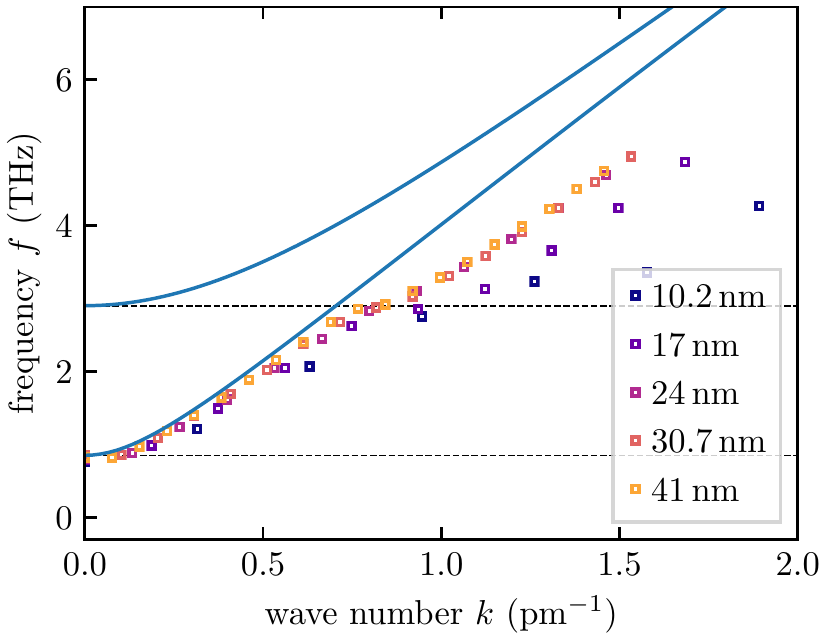}
    \caption{Spin-wave dispersion relation for Mn$_2$Au. Solid lines are numerical results obtained using \ac{LSWT}. Dotted lines are analytical predictions for the two positive \ac{AFMR} frequencies $f_0^\mathrm{a}$ and $f_0^\mathrm{b}$. The symbols represent the dispersion relation obtained for various Mn$_2$Au thicknesses using the method described in Ref.~\cite{Ritzmann2020}.}
    \label{fig:sw_dispersion}
\end{figure}

For the evaluation of the full spin-wave dispersion we use numerical methods that are explained in detail in the literature,  e.g.\ in Ref.~\cite{Donges2020}. The two branches of the spin-wave dispersion with positive frequencies close to the Brillouin zone center are shown in Fig.~\ref{fig:sw_dispersion} as solid lines. The two dotted lines are placed at the analytical predictions for the \ac{AFMR} frequencies $f_0^\mathrm{a}$ and $f_0^\mathrm{b}$ calculated above. 

In Ref.~\cite{Ritzmann2020}, a different approach was used to obtain the spin-wave dispersion relation for Fe. There, the authors also calculated the frequency spectra of spin waves excited by femtosecond spin-transfer torques for varying thicknesses. Using the condition for standing spin waves, i.e., $k_n=n\frac{\pi}{d}$, the authors were able to unambiguously relate the $n$-th peak in the spectrum with the wave number $k_n$. The spin-wave dispersion for Fe constructed in this way agrees well with analytical predictions based on \ac{LSWT}. If we follow the same procedure for Mn$_2$Au, we obtain what is shown as symbols in Fig.~\ref{fig:sw_dispersion}. While for frequencies below $f_0^\mathrm{b}$ the dispersion relation obtained this way agrees reasonably well with the results of \ac{LSWT} -- the small deviations could be a result of finite size effects since they are smaller for larger thicknesses -- it is completely off for $f\gtrsim f_0^\mathrm{b}$. The reason for this is that above $f_0^\mathrm{b}$ the peaks in the frequency spectrum could belong to either of the two branches. Moreover, due to the finite linewidth multiple peaks can overlap, possibly obstructing the unambiguous identification of individual peaks.

\clearpage

\section{Influence of FWHM and penetration depth on switching}
All results presented in the main text are obtained from simulations with a FWHM of the laser pulse of $\SI{40}{\femto\second}$ and a penetration depth of the spin current of $\lambda_\mathrm{STT}=\SI{1}{\nano\metre}$. Hereinafter, we vary these two parameters to investigate their impact on the switching of the Mn$_2$Au layer.

\begin{figure}[h]
    \centering
    \includegraphics[width=0.35\textwidth]{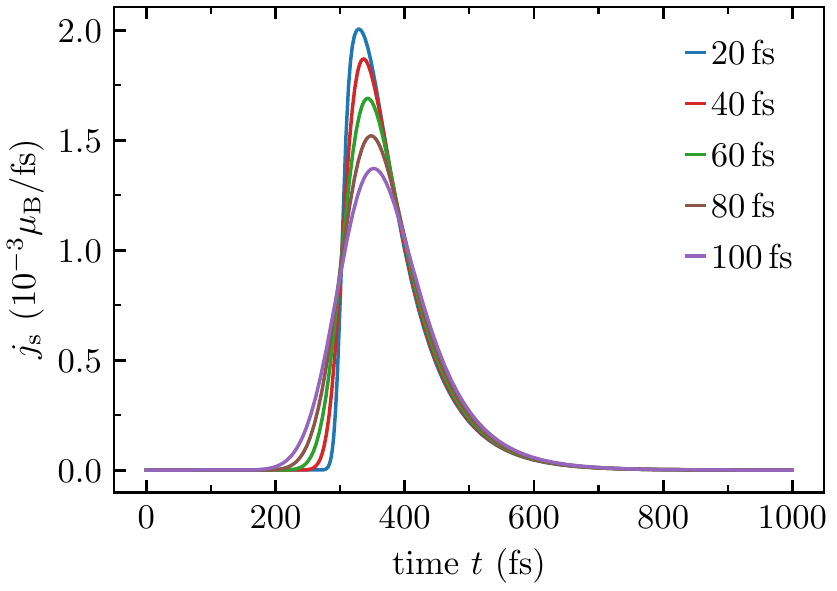}\, a)
\includegraphics[width=0.35\textwidth]{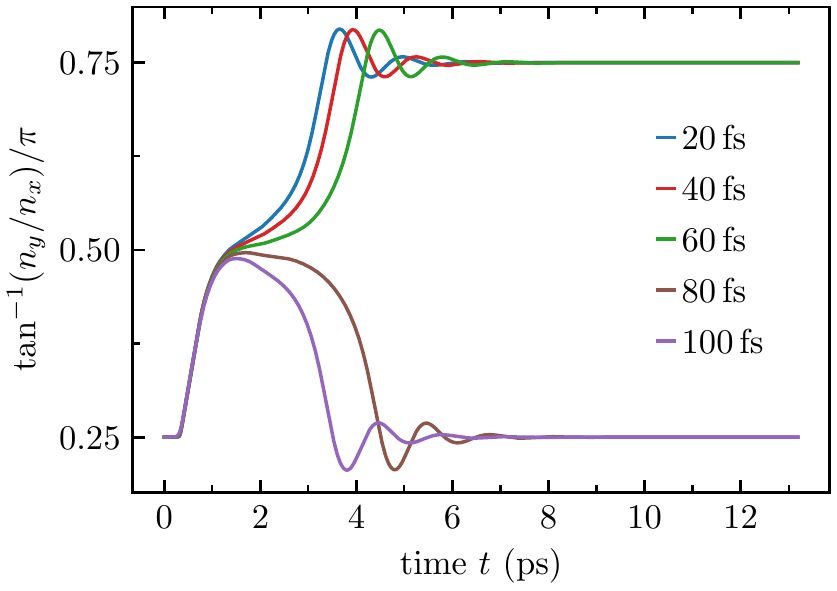}\, b)
\includegraphics[width=0.75\textwidth]{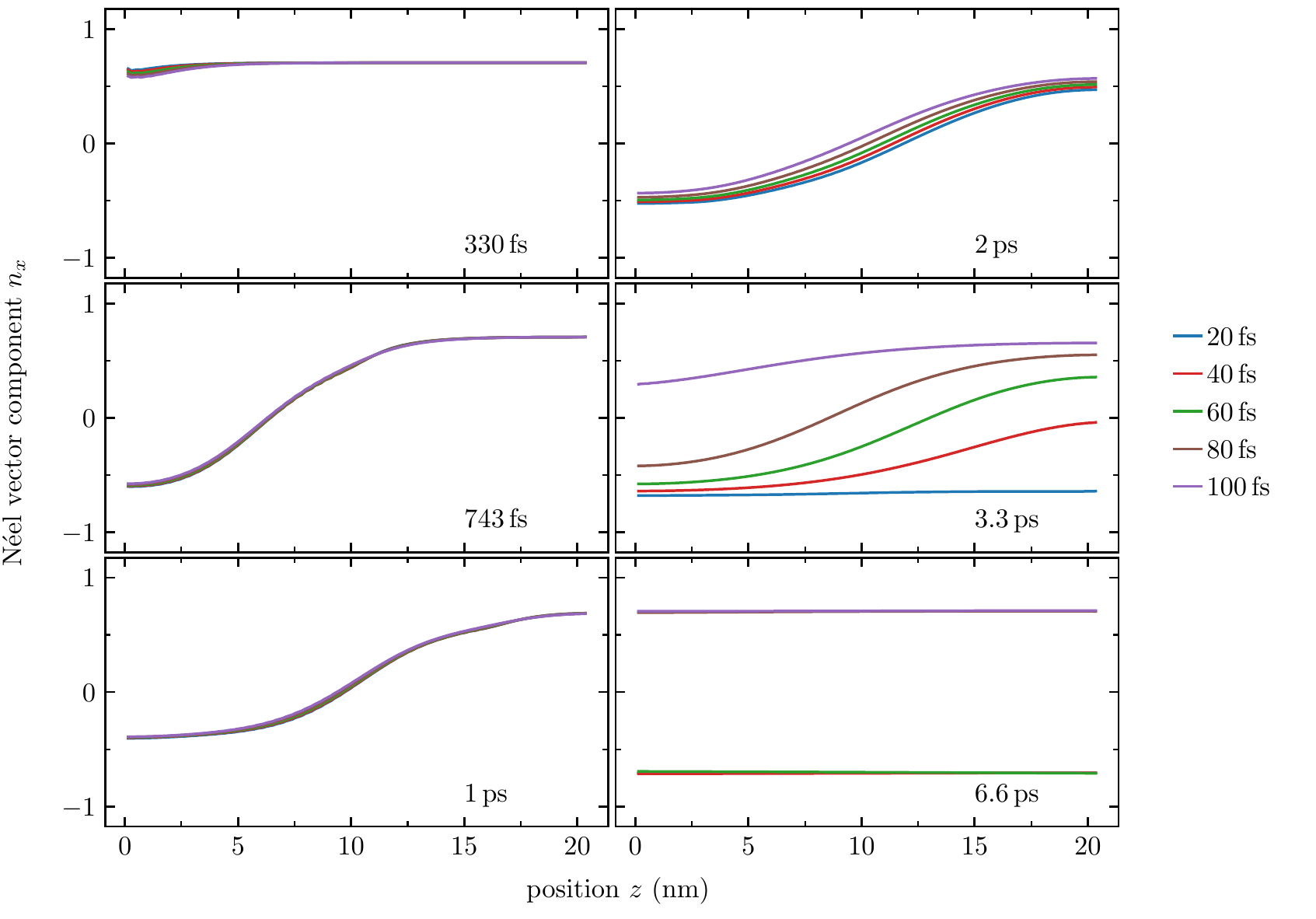}\, c) \\[-0.2cm]
    \caption{Impact of the width of the laser pulse on the switching process. a) Emerging spin currents at the Cu/Mn$_2$Au interface for different FWHMs of the Gaussian laser pulse. b) Time evolution of in-plane angle of the average N{\'e}el vector of Mn$_2$Au during the laser excitation. c)  N{\'e}el vector components versus distance from the interface to Cu (located at $z=0$) at different points in time.}
    \label{fig:neel_order_vs_time_and_FWHM}
\end{figure}

Varying the width of the laser pulse causing the demagnetization of the Fe layer for a fixed total laser fluence leads to a broadening of the temporal spin-current profile at the interface between Cu and Mn$_2$Au, see Fig.~\ref{fig:neel_order_vs_time_and_FWHM}a, while the total spin current stays constant. Consequently, the torque associated with the absorption of the spin-current, $\frac{j_\mathrm{s}(z,t)}{\mu_\mathrm{s}}\vec{S}_i\times(\vec{S}_i\times\hat{\vec{z}})$, persists for a longer time, while its maximum value is reduced. 

We find that for points in parameter space (the parameters are laser fluence, thickness, penetration depth, FWHM and Gilbert damping), which in the switching diagram are close to the border separating the regions of switching and no switching (cf.\ Fig.~5 of the main text), the FWHM can play a decisive role in the switching process.

E.g., for a laser fluence of $\SI{15.14}{\milli\joule/\cm\squared}$, a thickness of $d=\SI{20.5}{\nano\metre}$, a penetration depth of $\lambda_\mathrm{STT}=\SI{1}{\nano\metre}$ and a Gilbert damping parameter of $\alpha=0.01$, laser pulses with FWHMs of $\SI{20}{\femto\second}$ to $\SI{60}{\femto\second}$ lead to $90^\circ$ switching, in contrast to pulses with FWHMs of $\SI{80}{\femto\second}$ and $\SI{100}{\femto\second}$. This is demonstrated in Fig.~\ref{fig:neel_order_vs_time_and_FWHM}b, where the time evolution of the in-plane angle $\phi$ of the average N{\'e}el vector of Mn$_2$Au during the laser excitation is shown, given as $\phi = \tan^{-1}(n_y/n_x)\pi^{-1}$, in radians. $\phi = 0.75$ for a switching event and 0.25 for a non-switching event. In Fig.~\ref{fig:neel_order_vs_time_and_FWHM}c, the respective cross sections of the N{\'e}el vectors are shown at different points in time. 
\clearpage

\begin{figure}[h]
    \centering
        \includegraphics[width=0.35\textwidth]{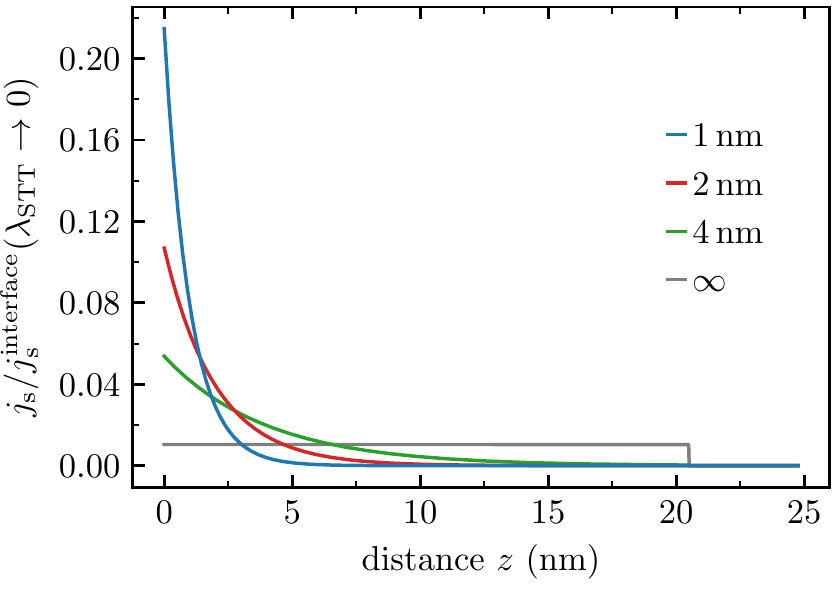}\, a)
    \includegraphics[width=0.35\textwidth]{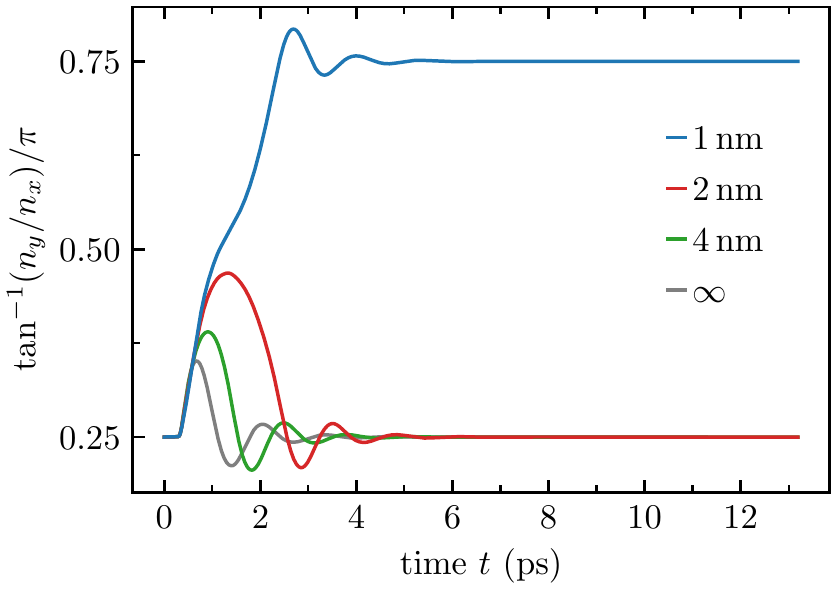}\, b)
\includegraphics[width=0.75\textwidth]{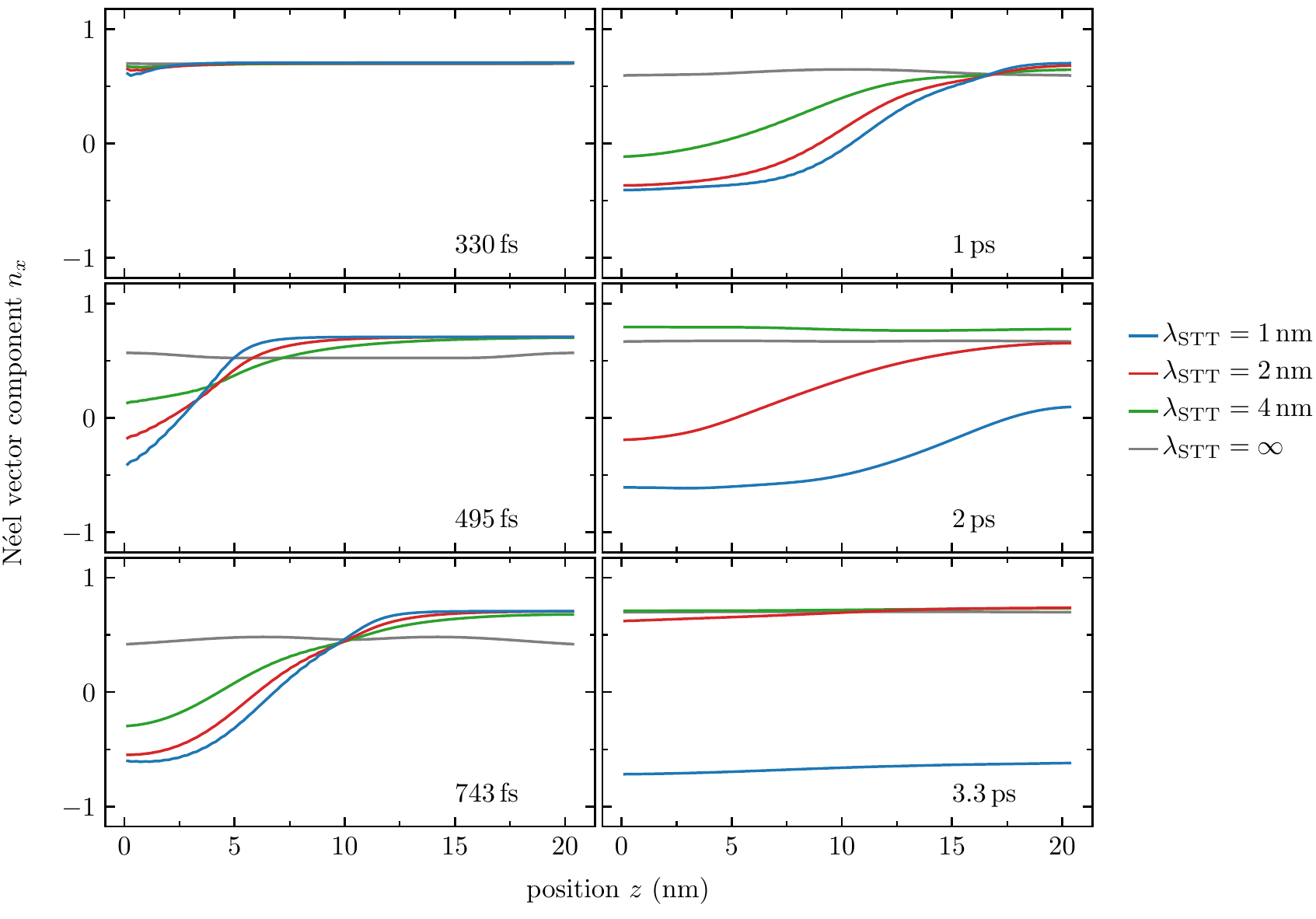}\,c) \\[-0.2cm]
\caption{Impact of the penetration depth $\lambda_\mathrm{STT}$ of the spin current on the switching process. a) Spatial profile of the spin current normalized to the spin current at the Cu/Mn$_2$Au interface for different $\lambda_\mathrm{STT}$. b) Time evolution of in-plane angle of the average N{\'e}el vector of Mn$_2$Au during the laser excitation. c)  N{\'e}el vector components versus distance from the interface to Cu (located at $z=0$) at different points in time.}
    \label{fig:cs_multiple_PDs}
\end{figure}

Next, we investigate the impact of the penetration depth $\lambda_\mathrm{STT}$ on the switching process.
Fig.~\ref{fig:cs_multiple_PDs}a shows the spatial profile the spin current for different penetration depths. Note that the spin current is normalized to the current that is transmitted at the interface between Cu and Mn$_2$Au. Henceforth, the area under the curves is independent of $\lambda_\mathrm{STT}$. The limit $\lambda_\mathrm{STT}\rightarrow \infty$ corresponds to a spatially homogeneous spin current/torque. For a laser fluence of $\SI{14.86}{\milli\joule/\cm\squared}$, a thickness of $d=\SI{20.5}{\nano\metre}$, a FWHM of $\SI{40}{\femto\second}$ and a Gilbert damping parameter of $\alpha=0.01$ only the spin current with $\lambda_\mathrm{STT}=\SI{1}{\nano\metre}$ leads to switching (see Fig.~\ref{fig:cs_multiple_PDs}b and c). We find that the excited spin dynamics become weaker for larger penetration depths, i.e.\ for equal total spin current but less localized absorption/torques. Thus, the emergence of switching of Mn$_2$Au is crucially related to the strongly peaked behavior of the spin-transfer torque.

\clearpage

\section{Impact of temperature on switching}
In the main text, we neglect the impact of temperature on the femtosecond-STT-induced spin dynamics in Mn$_2$Au. In an earlier work on Mn$_2$Au, thermal activation was identified as crucial for the switching via N{\'e}el-spin-orbit torques \cite{Selzer2022}. The impact of thermal fluctuations is twofold: first, it reduces the energy barrier, since it leads to a decrease of the effective anisotropy \cite{Callen1966}. Second, it renders the switching process stochastic. If thermal energies are comparable to the energy barrier, i.e.\ $k_\mathrm{B}T\gtrsim  d_{xy}$, and still smaller than $T_\mathrm{c}$, we expect quasi-superparamagnetic behavior of the N{\'e}el vector.

Spin dynamics at finite temperatures can be simulated using the stochastic Landau-Lifshitz-Gilbert equation,
\begin{align}
    \begin{split}
    \frac{\partial \vec{S}_i}{\partial t}
    =
    &-
    \frac{\gamma}{\mu_\mathrm{s}}
    \vec{S}_i
    \times
    \Big(
    \vec{H}^\mathrm{eff}_i
    +
    \vec{\zeta}_i
    \Big)
    +
    \alpha
    \vec{S}_i
    \times
    \frac{\partial \vec{S}_i}{\partial t}
    +
    \frac{j_\mathrm{s}(z,t)}{\mu_\mathrm{s}}
    \vec{S}_i
    \times
    (
    \vec{S}_i
    \times
    \hat{
    \vec{z}
    }
    ),
    \end{split}
    \label{eq:sLLG}
\end{align}
where the effective field $\vec{H}^\mathrm{eff}_i$ is supplemented with a stochastic field $\vec{\zeta}_i$, accounting for thermal fluctuations \cite{Nowak2007}. This field has the properties 
\begin{align}
    \langle 
        \vec{\zeta}_i(t)
    \rangle
    =0
    \hspace*{2em}
    \mathrm{and}
    \hspace*{2em}
    \langle 
        \vec{\zeta}_i(t)
        (\vec{\zeta}_j(t'))^\mathrm{T}
    \rangle
    =
    2
    \alpha
    k_\mathrm{B}T
    \frac{\mu_\mathrm{s}}{\gamma}
    \mathbbm{1}
    \delta_{ij}
    \delta(t-t').
\end{align}

\begin{figure}[h]
    \centering
\includegraphics[width=1.0\textwidth]{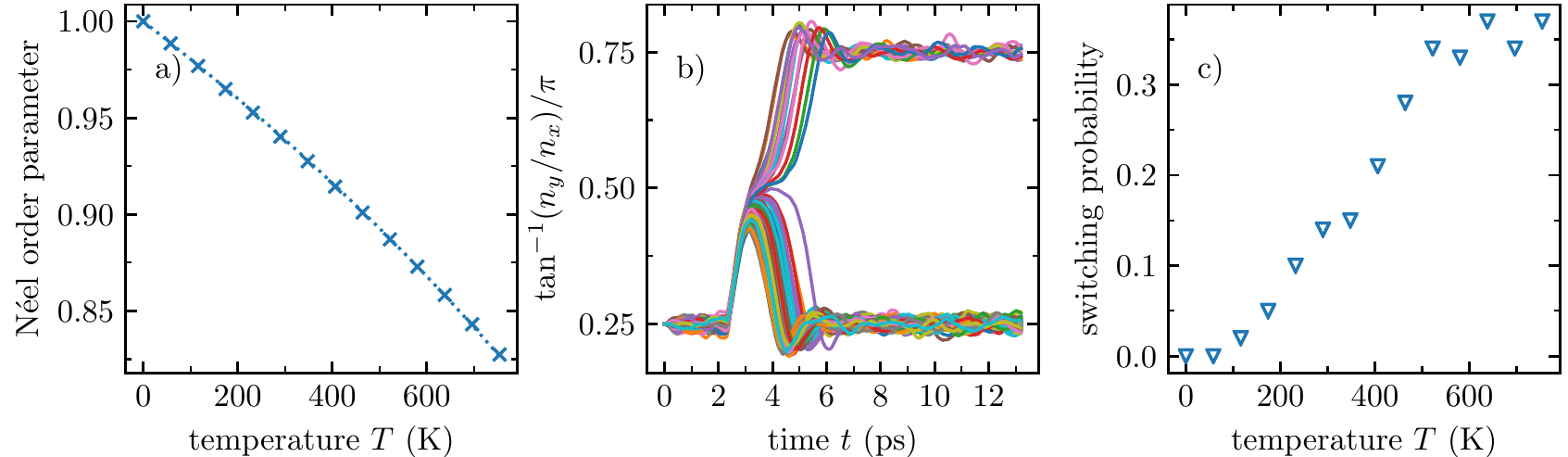} 
    \caption{Magnetization dynamics of Mn$_2$Au at finite temperatures. a) N{\'e}el order parameter versus temperature for a system with $\SI{20.3}{\nano\metre}\times \SI{20.3}{\nano\metre} \times \SI{20.5}{\nano\metre}$ and periodic boundary conditions in $x$- and $y$-direction and open boundary conditions in $z$-direction. b) 100 trajectories of the in-plane angle of the average N{\'e}el vector of Mn$_2$Au during the laser excitation at $T=\SI{290.1}{\kelvin}$. c) Probability for the switching from $[110]$ to $[\bar{1}10]$ versus temperature. We have used a laser fluence of $\SI{12.97}{\milli\joule/\cm\squared}$, a FWHM of $\SI{40}{\femto\second}$, a penetration depth of $\lambda_\mathrm{STT}=\SI{1}{\nano\metre}$ and a Gilbert damping parameter of $\alpha=0.01$ in b) and c).}
    \label{fig:switching_vs_T}
\end{figure}

At finite temperatures, the N{\'e}el order parameter, the absolute value of the temporal and spatial average of the N{\'e}el vector, is reduced, see Fig.~\ref{fig:switching_vs_T}a. The dotted line corresponds to the theoretical prediction for a classical magnet, $n(T)/n(0)=(1-\frac{T}{T_\mathrm{c}})^{0.33}$, with $T_\mathrm{c}=\SI{1720}{\kelvin}$. Note that the $T_\mathrm{c}$ for the system investigated here is slightly larger than the value of $\SI{1680}{\kelvin}$ that was calculated in Ref.~\cite{Selzer2022}. This is due to finite-size effects and the fact that we use periodic boundary conditions in $x$- and $y$-direction.

Fig.~\ref{fig:switching_vs_T}b displays the time evolution of the order parameter during a laser excitation close to room temperature ($T=\SI{290.1}{\kelvin}$). In contrast to zero temperature, the system switches in a finite number of simulations. This is analyzed in more detail in Fig.~\ref{fig:switching_vs_T}c. There, we demonstrate an increase of the switching probability with temperature in the range of $T \in [0\si{\kelvin},755\si{\kelvin}]$, with indications of convergence at high temperatures. In contrast to what is reported in Ref.~\cite{Selzer2022}, the switching probability appears to converge to a value far from $100\%$. We attribute this to the fact that the torque associated with the process investigated here occurs on a much faster timescale: in Ref.~\cite{Selzer2022}, the system gets pushed to a state close to the maximum of the switching barrier for a couple of picoseconds due to the effect of the N{\'e}el-spin-orbit torques. During that time period, there is a finite chance of overcoming the remaining energy difference to the maximum of the switching barrier by thermal activation. Once the system has overcome the barrier, i.e.\ it has switched, the chance of switching back to the initial state is very small. Since the N{\'e}el-spin-orbit torques are active over a comparably long timescale ($\sim 10$ ps), the chance of ending up in the switched state becomes close to $100\%$ for higher temperatures. Here, the spin-current pulse that induces the torques lasts only a couple of hundred femtoseconds. Thus, the time interval, within which the remaining energy barrier can be overcome by thermal activation, is much shorter, leading in general to a less pronounced impact of thermal fluctuations on the switching dynamics.

%